\documentclass[]{aa}  

\usepackage{amsmath}
\usepackage{amssymb}
\usepackage{color}
\usepackage{url}
\usepackage{ulem}
\usepackage{multirow}
\usepackage{graphics}
\usepackage{longtable}
\usepackage{graphicx}
\usepackage{bm}
\usepackage{appendix}

\usepackage{txfonts}

\DeclareGraphicsExtensions{.pdf,.jpeg,.png,.jpg,.pdf}

\usepackage{wasysym}

\usepackage{twoopt}
\usepackage[breaklinks=true,pagebackref=false,colorlinks=true,citecolor=blue]{hyperref} 
\bibpunct{(}{)}{;}{a}{}{,} 

\makeatletter
\renewcommand*\aa@pageof{, page~\thepage{} of~\pageref*{LastPage}}
 
\makeatother
\newcommand{\cdc}{\color{black}}

\begin{document} 

 \titlerunning{Constraints from the orbital period ratios}
 \authorrunning{Chen, Mordasini, Xie, Zhou \& Emsenhuber}
   \title{Constraints on the formation history and composition of Kepler  planets from their distribution of orbital period ratios}

   \author{Di-Chang Chen
          \inst{1,2,3,*}, 
          Christoph Mordasini\inst{3}, Ji-Wei Xie\inst{1,2}, Ji-Lin Zhou\inst{1,2} \and Alexandre Emsenhuber\inst{4}
          }
   \institute{School of Astronomy and Space Science, Nanjing University, Nanjing 210023, China\\
              \email{dcchen@nju.edu.cn}
   \and
   Key Laboratory of Modern Astronomy and Astrophysics, Ministry of Education, Nanjing 210023, China
   \and
   Weltraumforschung und Planetologie, Physikalisches Institut, Universität Bern, Gesellschaftsstrasse 6, 3012 Bern, Switzerland
   \and
   Universitäts-Sternwarte, Ludwig-Maximilians-Universität München, Scheinerstraße 1, 81679 München, Germany\\
   \renewcommand{\thefootnote}{}       
   \footnotetext{* LAMOST Fellow}}
   \date{Received ; accepted}
 
  \abstract
   {The Kepler high-precision planetary sample has revealed a `radius valley' (a paucity of planets of $\sim$$1.9 R_\oplus$), separating compact super-Earths from sub-Neptunes with lower density.
   Super-Earths are generally assumed to be rocky planets that were probably born in-situ, while the composition and origin of sub-Neptunes remains debated (in-situ rocky cores with H/He versus migrated ex-situ water-rich planets). Numerous statistical studies have explored planetary and stellar properties (e.g. mass, age, metallicity) and their correlations to provide observational clues. However, { no conclusive result on the origin of radius valley or the composition of sub-Neptunes has been derived to date.} 
   }
   {To provide more constraints on the formation history and composition, we aim to investigate the orbital spacing of sub-Neptunes and super-Earth planets in multiple transiting systems and compare with theoretical predictions of planet pairs of different formation pathways and compositions in synthetic planetary systems.}
   {Based on the planetary sample of Kepler multiple planet systems, we derive the distributions of orbital period ratios of sub-Neptune and super-Earth planet pairs and calculate the normalised fraction of near-first-order mean motion resonances.
   Using  synthetic planetary systems generated by the Generation III Bern Model, we also obtain theoretical predictions of period ratio distributions of planet pairs of different compositions and origins: { water-rich sub-Neptunian planets that was born ex-situ and then migrated inward versus water-poor super-Earth that were born inside of the ice line.}
   }
   {{ We find that actual Kepler sub-Neptune pairs show a significant preference to be near-first-order MMR by a factor of $1.7^{+0.3}_{-0.3}$. 
   This is smaller than the model predictions for `water-rich' pairs, but larger than that of `water-poor' pairs by confidence levels of $\sim 2-\sigma$.}
   Actual Kepler Super-Earth pairs show no significant difference from a random distribution without a significant preference for MMR. The derived normalised fraction of near-first-order-resonances { of actual Kepler Super-Earth pairs} is generally consistent with that of `water-poor' model planet pairs within $1-2 \sigma$ uncertainties but significantly ($\gtrsim 3-\sigma$) smaller than that of synthetic `water-rich' planet pairs.}
   {Based on the distributions of orbital period ratios, we conclude that orbital migration has been more important for sub-Neptunes than for super-Earths, suggesting a partial ex situ formation of the former and an origin of the radius valley caused in part by distinct formation pathways. However, the model comparisons also show that sub-Neptunes in actual Kepler multiple systems are not likely ($\sim 2 \sigma$) to be all water-rich/ex situ planets but a mixture of the two (in situ/ex situ) pathways. Whereas, Kepler super-Earth planets are predominantly composed by of water-poor planets that were born inside the ice line, likely through a series of giant impacts without large scale migration.}


   \maketitle

\section{Introduction} 
\label{sec:intro}
Thanks to the improvement of technology and long-term surveys, the number of known planets has exceeded 5,000 and thousands of candidates are yet to be confirmed \citep[NASA Exoplanet Archive;][]{2013PASP..125..989A}.
Among these surveys, the Kepler mission has played an important role (e.g. nearly half of identified planets) and provided an unprecedented high-precision sample for exoplanet science \citep{2011AJ....142..112B,2014ApJS..211....2H,2017ApJS..229...30M}.
The majority of planets (candidates) detected by Kepler are compact super-Earths ($\lesssim 1.7 R_\oplus$) and sub-Neptunes ($\gtrsim 2.1 R_\oplus$) with lower bulk densities, separated by a `Radius valley' \citep[$\sim 1.9 \pm 0.2 R_\oplus$; e.g.][]{2017AJ....154..109F,2018MNRAS.480.2206O,2018MNRAS.479.4786V,2023ASPC..534..863W}. 
Super-Earths are generally assumed to be `water-poor' rocky planets (iron and silicates with no/thin gas envelope) which are probably formed in-situ \citep[e.g.][]{2018NatAs...2..393S,2019NatSR...911683B,2020A&A...636A..58A}.

{ However, the composition of sub-Neptunes (as well as the origin of radius valley) remains debated.}
Plenty of theoretical models have proposed, which can be generally classified into two categories.
Some studies suggest that sub-Neptunes are born as `gas dwarfs' consisting of rocky cores plus a  H/He-dominated gas envelope massive enough (i.e. larger a few percent in mass) to significantly increase the radius.
During the subsequent evolution, some of these planets lost their gas envelope and evolved to super-Earths, resulting in the appearance of radius valley.
The energy source that drives the gas loss process could be either from the radiation of the host star \citep[the photo-evaporation;][]{2013ApJ...775..105O,2014ApJ...795...65J,2016ApJ...818....4L,2017ApJ...847...29O,2018ApJ...853..163J}  or from the cooling power of the planet core \citep[the core-powered model;][]{2016ApJ...825...29G,2018MNRAS.476..759G,2019MNRAS.487...24G,2020MNRAS.493..792G}.
Alternatively, some other studies propose that sub-Neptunes are `water-worlds' containing significant amounts (several tens of percent of their total masses) of $\rm H_2 O-$dominated fluid/ice  \citep{2019PNAS..116.9723Z,2020ApJ...896L..22M,2021ApJ...914...84A}.
To obtain such large water contents, these `water-rich' sub-Neptunes are expected to have formed beyond the ice line and migrated inward to current orbits due to mechanisms such as planet-disk interactions \citep{1997Icar..126..261W,2013ApJ...775...42I,2019A&A...624A.109B,2020A&A...643L...1V,2020A&A...644A.174V,2022ApJ...939L..19I,2024NatAs.tmp...33B}.
\footnote{ It is worth noting that the water-rich and water-poor planets are classified based on their formation characteristics (e.g. composition, location, etc.), whereas super-Earth and sub-Neptune planets are classified based on their current physical radius (i.e. after formation and any evolution that occurred such as migration or atmospheric loss).}

To reveal the origin of the radius valley and the composition of sub-Neptunes, numerous of studies have investigated the planetary mass-radius curves and the dependence of the radius valley on planetary and stellar properties \citep[e.g. orbital period, stellar mass, metallicity, age;][]{2018MNRAS.479.4786V,2018MNRAS.480.2206O,2020AJ....160..108B,2022AJ....163..249C}.
However, the current observational data provides no conclusive result to distinguish these theoretical hypotheses as none of them could explain all the observational evidences.
For example, \cite{2022AJ....163..249C} shows that the average radii of sub-Neptunes shrink with age using the LAMOST-Gaia-Kepler kinematic catalogue \citep{2021ApJ...909..115C,2021AJ....162..100C}, supporting that some sub-Neptunes are `gas-dwarfs' containing significant gas envelope.
Whereas, some other studies provide supporting observational evidences that many sub-Neptunes are `water worlds' from the planetary mass-radius curves \citep{2019PNAS..116.9723Z} and the planetary radius distribution around M-type stars \citep{2022Sci...377.1211L}.

Besides atmospheric spectroscopy \citep[e.g.][]{2023ApJ...953...57K},
more clues on the formation and composition of sub-Neptunes can be derived from their orbital spacing (the period ratio of adjacent planets) in multiple planetary systems.
Specifically, if sub-Neptunes are `water-rich', they are thought to have formed beyond ice lines and then migrated inward to current orbits.
Convergent migration in gas-damped disk can lead to capture in mean motion resonances (MMR), { such that ‘water-rich’ sub-Neptunes would more likely exhibit close spacing and binding in MMRs} \citep[e.g.][]{2001A&A...374.1092S,2002ApJ...567..596L,2008A&A...483..633P,2021A&A...656A..69E}. { Thus, if sub-Neptunes are formed ex-situ, sub-Neptune pairs would have a larger fraction in MMR comparing to a random distribution.}
On the contrary, if sub-Neptunes and super-Earths had the same formation pathway and just differ by their subsequent evolution (rocky cores that have kept/lost their thick gas envelopes), they were probably formed both more or less in-situ via solid accretion and/or giant impacts of embryos due to gravitational interactions  \citep[e.g.][]{2001Icar..152..205C,2002Natur.418..949Y,2007Natur.450.1206T,2007ApJ...671.2082K,2021A&A...656A..69E}.
In this case, we would not expect distinct frequencies of MMRs from a random distribution.  
{ The subsequent long-term gravitational interactions would lead to the widening of the orbital spacing and breaking of some unstable MMRs \citep[e.g.][]{2007ApJ...666..423Z,2012ApJ...756L..11L,2015ApJ...807...44P,2019NatAs...3..424M}. 
Some recent studies have also suggested that atmospheric escape would contribute to the breaking of MMRs \citep{2020ApJ...893...43M,2023AJ....165..174W}.
Thus, generally, planets formed via in-situ pathway are less likely to be in MMRs because of these above mechanisms.}
Therefore, by exploring the distributions of orbital period ratio of sub-Neptune pairs versus super-Earth pairs, one can put constraints on their origin and the compositions: what are the predominant compositions of sub-Neptunes (rocky+H/He versus water-rich) and how much do these two different compositions and thus formation pathways contribute?



In this paper, by using the Kepler data and a synthetic population generated by the Bern Model of Planet Formation and Evolution, we investigate the distribution of period ratios of adjacent planet pairs. 
The rest of this paper is organised as follows.  
{ In Sect. \ref{sec:sample}, we describe how we construct our observed and synthetic planetary samples.}
In Sect. \ref{sec.PR.obs}, by analysing the observational sample selected from Kepler DR25, we present the observational evidences that sub-Neptunes show a preference to be captured in near-MMR. 
In Sect. \ref{sec.PR.theory}, we obtain the theoretical predictions of period ratio distributions for the planet pairs with different compositions by using the synthetic planet population.
In Sect. \ref{sec.dis.implication}, we put constraint on the formation and composition of planets by comparing the observational evidences with  the theoretical results. 
Finally, we summarise the paper in Sect. \ref{sec.Summary}.

\section{Sample} 
\label{sec:sample}
In this section, we describe how we construct the planetary samples from the observed Kepler data and from a synthetic population generated by the Bern model. 

\subsection{Observational sample}
For the observational sample, we initialised on the Kepler data release (DR) 25, which contains 8,054 Kepler objects of interest (KOI) \citep{2017ApJS..229...30M}.
Here, we excluded KOIs flagged as false positives (FAP), leaving 4034 planets (candidates).
Since the synthetic systems are generated around single $1 M_\odot$ stars, we only keep systems around Sun-like stars (i.e. effective temperature in the range of $4700-6500$ K and surface gravity $\log g>4$).
We also removed potential binaries by excluding planet host stars with Gaia DR2 re-normalised unit-weight error (RUWE) $>1.2$.
For the planetary sample, we adopt the following criteria:
\begin{enumerate}
    \item Excluding the Kepler Objects of Interest (KOI) identified as false positives;
    \item Removing KOIs with radii $>4 R_\oplus$;
    \item Removing KOIs with period $>400$ days to ensure the detection efficiency;
    \item Excluding KOIs with period $<5$ days to avoid the influence of tide \citep[e.g.][]{2008ApJ...678.1396J,2012MNRAS.423..486L};
    \item Keeping planetary systems with multiplicity $N_{\rm p} \ge 2$.
\end{enumerate}
{ Planets (candidates) with radii $\ge 2.1 R_\oplus$, $\le 1.7 R_\oplus$, and $1.7-2.1 R_\oplus$ are classified as sub-Neptunes, super-Earths, and valley planets, respectively.
The final observational sample contains 697 planets/candidates (373 sub-Neptunes, 239 super-Earths and 85 valley planets) in 293 multiple systems.}

\begin{figure}[!t]
\centering
\includegraphics[width=\linewidth]{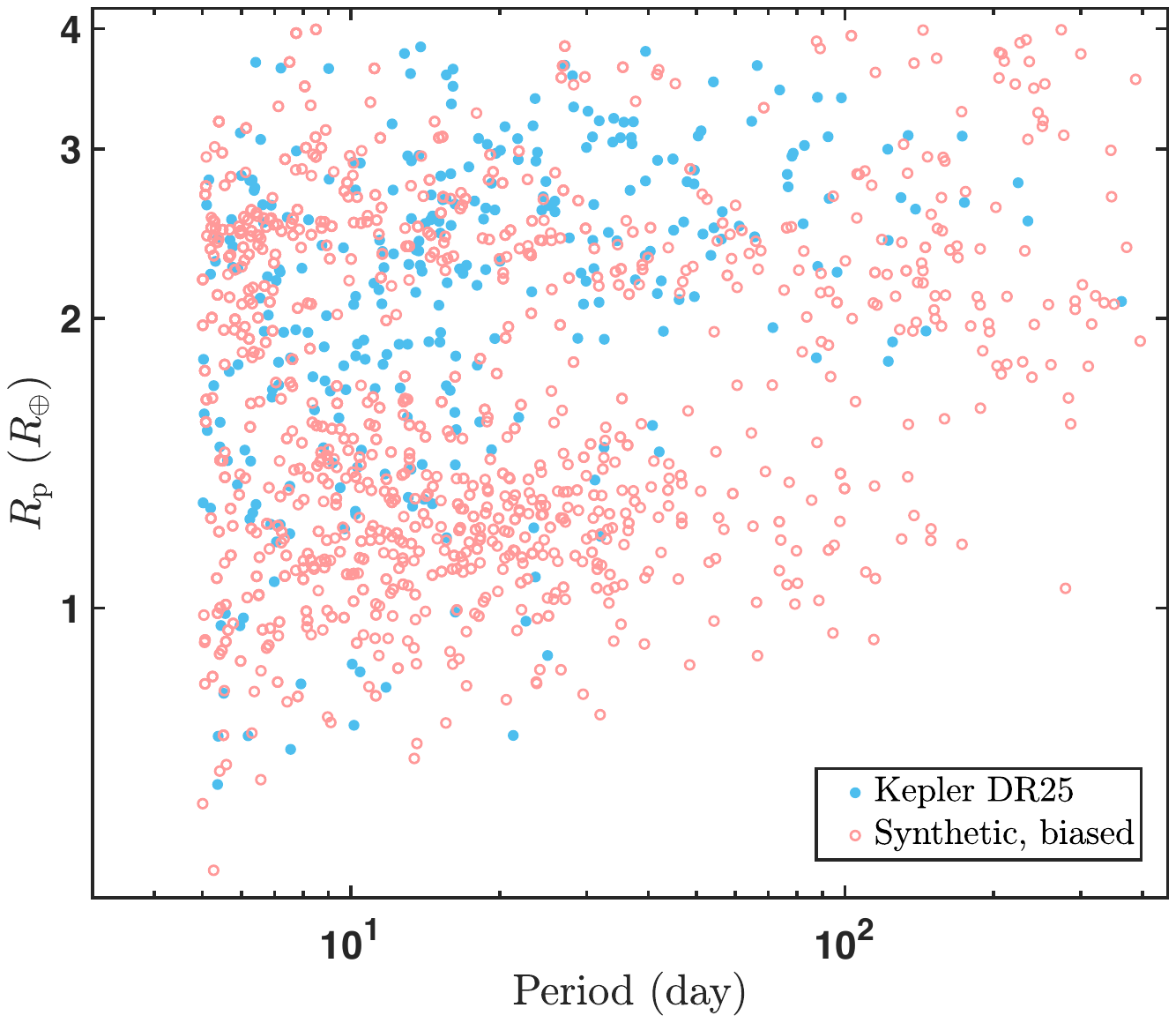}
\caption{Radii as a function of orbital period of planets selected from Kepler DR 25 data (blue) and the synthetic population after applying the bias of the Kepler survey (pink).
We restricted the sample to planets with radii between $1-4 R_\oplus$ and orbital periods between $5-400$ days in multiple systems.
\label{figradiusperioddigram}}
\end{figure}

\subsection{Synthetic sample}
We adopt a combined planetary formation and evolution model (the Generation \uppercase\expandafter{\romannumeral3} Bern model) to synthesise a population of model planetary systems.
The Bern model \citep{2004A&A...417L..25A,2005A&A...434..343A,2013A&A...558A.109A,2009A&A...501.1139M,2012A&A...547A.112M,2021A&A...656A..69E} simulates the initial formation and subsequent long-term evolution phases of planetary systems in a self-consistently coupled way. A detailed description of the Generation \uppercase\expandafter{\romannumeral3} Model used here can be found in \cite{2021A&A...656A..69E}; therefore we here only give a short overview. Starting from planetary embryos and background planetesimals embedded in a gaseous disk, the formation stage includes the following main physical processes: the viscous evolution of the protoplanetary gas disc including disk photoevaporation, the evolution of the surface density and dynamical state of planetesimals, the simultaneous accretion of planetesimals and gas by the embryos, the orbital migration and eccentricity and inclination damping of the protoplanets by the gas disk, and the gravitational N-body interactions  among the multiple protoplanets forming in each system including impacts. The following evolution stage follows the long-term evolution of each planet individually. It includes the following processes: the thermodynamic evolution (cooling and contraction of the interior), mass loss via atmospheric escape, and tidal migration due to stellar tides. The individual sub-modules included in the model can be of a considerable complexity and whenever possible, underlying differential equations are solved rather than using semi-empirical closed-form expressions. However, in order to still be able to simulate many systems and planets over Gigayears, numerically a low-dimensional approach is used (except for the N-body integrator which is 3D): the gas disk is assumed to be rotationally symmetric, the planet interiors are spherically symmetric and the planetesimals are described on a 1D grid (radial) by a surface density rather than by individual particles. 

For the analysis, we use an improved version of the nominal synthetic population for solar-mass stars (\texttt{NG76}) that was presented in \cite{2021A&A...656A..70E}. The improvements in the population studied here (called \texttt{NG76longshot}) regard two aspects (see \cite{2022arXiv220310076W}, \cite{2023A&A...673A..78E} and \cite{2024NatAs.tmp...33B} where this improved data set was also already analysed): firstly, the formation stage which includes \textit{N}-body interactions was prolonged from 20 to 100 Myr to capture late dynamical events \citep{2017MNRAS.470.1750I}. Secondly, during the evolutionary phase the originally employed simple energy-limited XUV-driven escape model \citep{2018ApJ...853..163J} was replaced with a full hydrodynamic model \citep{2018ApJ...866L..18K,2023arXiv230702566A} and the equation of state for water \citep[AQUA,][]{2020A&A...643A.105H} now includes the different physical phases. Furthermore, it is now assumed that the water mixes with H/He (where present) instead of a onion-like structure with separated layers. This has important implications for the radii of water-rich close-in sub-Neptunes because of the runaway greenhouse radius inflation effect \citep{2020A&A...638A..41T}. 

Despite its high content in physical processes, it should be noted that the model is based on classical concepts - more recent developments like a description of growth starting from dust over pebbles and planetesimals to protoplanets \citep{2020A&A...642A..75V,2022A&A...666A..90V}, hybrid pebble and planetesimal accretion \citep{2018NatAs...2..873A,2023A&A...674A.144K}, MHD-wind driven disk evolution \citep{2023A&A...674A.165W}, the impact of structured disks \citep{2022A&A...668A.170L,2023MNRAS.518.3877J} or a torque densities approach for orbital migration \citep{2022A&A...664A.138S}
 are not yet included.  These limitations should be kept in mind when assessing the results shown here.

For the population synthesis studied here (New Generation Planetary Population Synthesis, NGPPS), the 1000 systems were obtained by varying key disk initial conditions in a Monte-Carlo way \citep{2004ApJ...604..388I,2009A&A...501.1161M}. The probability distributions were derived from observations of protoplanetary disks \citep{2018ApJS..238...19T} and are described in detail in \cite{2021A&A...656A..70E}. Each disk is initially seeded with 100 lunar-mass embryos at t=0. The variation of the disk properties over the observed range leads to a large diversity of resulting planetary systems, from systems with only low-mass Super-Earths to multiple massive giants. 

From this  synthetic population, we select Kepler-like planets with the same criteria as the observational sample (i.e. planets with radii $\le 4 R_\oplus$ and period within $5-400$ days) at 5 Gyr.
In order to compare with Kepler observations, we then apply the synthetic detection bias using the KOBE program \citep[for more details, see Appendix C of][]{2021A&A...656A..74M}.
Specifically, we first use the KOBE-Shadow module to test whether generated planets could transit.
Then for these transiting planets, we use the KOBE-transit module to calculate their transit parameters and keep those with numbers of transit $\ge 3$ and signal-over-noise $\ge 7.1$.
Finally, we use the KOBE-Vetter module to simulate the detection completeness and reliability.
After applying the detection biased, only systems of two or more detected synthetic planets are retained.
{ The final synthetic sample consists of 1,851 planets (740 sub-Neptunes, 920 super-Earths and 191 valley planets) in 757 multiple systems.}
Figure \ref{figradiusperioddigram} shows the radius-period diagram of the selected observational and synthetic samples.

\section{Analysis of the selected Kepler sample}
\label{sec.PR.obs}

In this section, using the selected observational sample, we obtain the period ratios (PR) of adjacent planet pairs in multiple systems and investigate whether they have a larger fraction in near-MMRs compared to a random distributions.

\subsection{Constructing a control sample}
To evaluate whether the observational planets are randomly paired or not, we construct a control sample from the selected observational sample.
Specifically, for each planet { (sub-Neptune/super-Earth/valley planet)} in a given system, we randomly assign a orbital period from the observed distribution of planets in the select planetary sample.
We then calculate its signal-over-noise $SNR$ and transit detection efficiency.
If the resulting $SNR \ge 7.1$ and detection efficiency $\ge 10\%$, we consider it observable by Kepler and keep it.
Otherwise, we will reassign the orbital period to it and repeat the above process until the above criteria are met.
We them repeat the above procedure for 1,000 times and construct the control sample.
Finally, we calculate the period ratios for adjacent planet pairs in the control sample and compare with the observational results from Kepler data.


\subsection{Ensuring the stability of multiple systems}
{ To evaluate the stability of multiple systems in the observational control sample, we calculate the Hill-stability criterion from the orbital separation,
\begin{equation}
    H \equiv \frac{a_{\rm out}-a_{\rm in}}{R_{\rm H}},
\end{equation}
where the two planets are indexed as `in' and `out' and $a$ denotes the semi-major axis.
We here select planet pairs with $H > 7.1$ as Hill-stable when the distribution of period ratio for the control sample is most similar (with the largest KS $p-$value) to that of the observed sample (see Appendix \ref{sec.Appendix.Hillstable} for details).
Actually, the theoretical limit for Hill stability $K$ is dependent on various parameters, such as,  multiplicity $N_{\rm p}$ \citep{2010A&A...516A..82F}, planetary mass ratio over stellar mass $\mu$ \citep{1996Icar..119..261C,2007ApJ...666..423Z}, the orbital eccentricities/inclinations \citep{1999Icar..139..328Y,2007ApJ...666..423Z}, and stable timescales \citep{2009Icar..201..381S}.
The value of $K$ is $\sim 2\sqrt{3}$ to $12$ from simulation and observation \citep{1993Icar..106..247G,1996Icar..119..261C,2015ApJ...808...71M,2015ApJ...807...44P,2019MNRAS.490.4575H,2024AJ....167...46D}.
In Appendix \ref{sec.Appendix.Hillstable}, we make detailed discussions on the effect of the variation of $K$.}
Here we evaluate the Hill-stability by assuming adjacent planets on circular orbits (i.e. $e=0$) since most of Kepler planets have no (accurate) eccentricity measurements, and on average, the Kepler planets in multiple systems are on nearly circular orbits \citep[$e \sim 0.04$; e.g.][]{2016PNAS..11311431X}.
$R_{\rm H}$ is the mutual Hill radius relevant for dynamical interactions \citep{1982CeMec..26..311M}, given by
\begin{equation}
    R_{\rm H} = \left( \frac{M_{\rm in}+M_{\rm out}}{M_*} \right)^{1/3} \frac{(a_{\rm in}+a_{\rm in})}{2},
\end{equation}
where $M$ is the mass of the planets ($\rm in$ and $\rm out$) and star ($*$).
Since most planets in our observational control sample have no (accurate) mass measurements, we estimate their masses by converting their measured radii according to an empirical broken power-law mass–radius relationship \citep{2011Natur.470...53L,2012ApJ...750..148W,2020A&A...634A..43O,2017A&A...602A.101R,2023arXiv231112593B,2023MNRAS.522..828H}:
\begin{equation}
    \frac{M_{\rm p}}{M_\oplus} = C \left( \frac{R_{\rm p}}{R_\oplus} \right)^{\alpha},
\end{equation}
where $C$ and $\alpha$ are set as 3.98 and 0.92 for $R_{\rm p} \le 3.80 R_\oplus$ and 0.74 and 2.18 for $R>3.80 R_\oplus$, respectively \citep{2017A&A...602A.101R}. 
{ It is clear that this power-law only reflects a mean mass-radius relation. On the other hand, the dependency of the Hill sphere on mass is weak (proportional to only $M^{1/3}$). }    

For systems of three or more planets, the long-term interaction between planets could potentially induce additional instabilities \citep{1996Icar..119..261C,2009Icar..201..381S}.
To further ensure the stability of planet pairs in systems with $N_{\rm p} \ge 3$, we adopt a conservative heuristic criterion suggested by \cite{2014ApJ...790..146F}: 
\begin{equation}
    H_{\rm in} + H_{\rm out} > 18,
\end{equation}
where `in' and 'out' denote the inner pair and the outer pair of three adjacent planets.

\subsection{Comparing the period ratio distributions of the observational sample with control sample}



\begin{figure}[!t]
\centering
\includegraphics[width=1\linewidth]{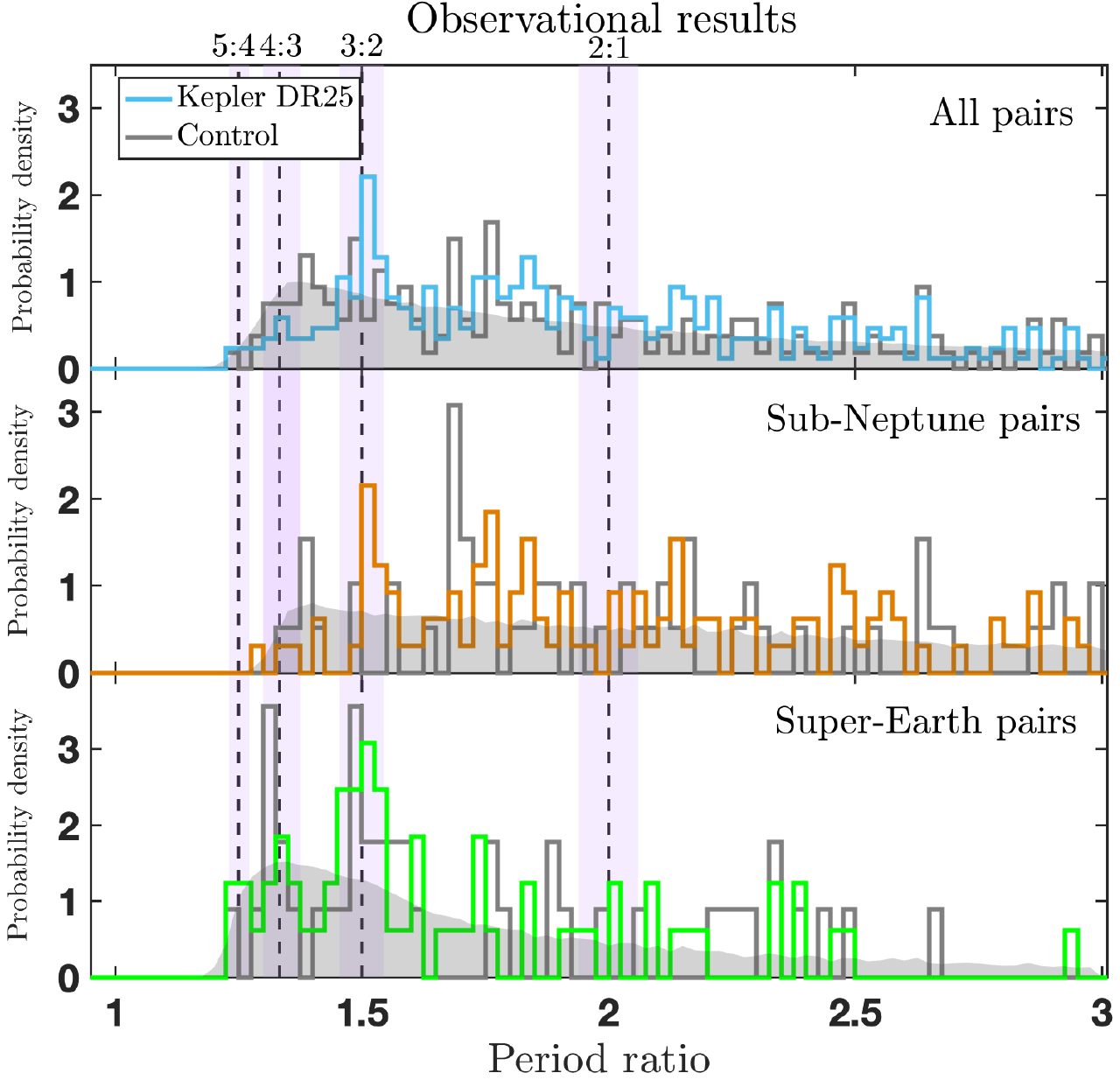}
\caption{ Probability density functions of period ratios of adjacent planet pairs derived from the Kepler DR 25 data (blue) and from a  control sample with the assumption that planets are randomly paired (grey region for the control sample by combining the 1,000 times of randomly pairing and grey line is one example with similar fractions in near-first-order MMRs in the 1,000 sets). 
The dotted lines and purple band represent the centres and regions of the first-order resonances.
\label{figPRhistobservation}}
\end{figure}


Figure \ref{figPRhistobservation}  displays the histogram distribution function of period ratio derived from the observational sample and control sample for all planet pairs (top panel), sub-Neptune pairs (i.e. pairs of two sub-Neptunes; middle panel), and Super-Earth pairs (i.e. pairs of two super-Earths; bottom panel).
As can be seen, actual Kepler planet pairs seem to prefer to be near first-order MMR (i.e. 5:4, 4:3, 3:2 and 2:1) compared to the random paired control sample.
Furthermore, sub-Neptune pairs show an obvious preference in near-MMR comparing to the corresponding control sample, which is in contrast not seen in Super-Earth pairs.
To describe the above feature mathematically, according to previous studies \citep[e.g.][]{2020AJ....160..180J,2023MNRAS.522..828H}, 
we select planet pairs with PRs between $[\frac{3j+4}{3j+1}, \frac{3j+2}{3j-1}]$ (the closest third-order resonances) and $|\Delta| < 0.03$ are selected as near-MMR, where $\Delta$ is a dimensionless parameter to measure the offset of a given PR to $j+1:j$ MMR, 
\begin{equation}
    \Delta \equiv \frac{j}{j+1} {\rm PR}-1.
\end{equation}
We then calculate the fractions of planet pairs being near-MMRs for both the actual observational sample $F^{\rm obs}_{\rm MMR}$ and the random paired control sample $F^{\rm con}_{\rm MMR}$.
We here adopt 0.03 as a typical value for the boundary of near-MMRs. Actually, the expected value of the boundary depends on the planetary masses and eccentricities and is $\sim$ several of $10^{-2}$ both from Numerical simulations \citep[e.g.][]{2015MNRAS.453.4089S,2014ApJ...786..153X,2015MNRAS.453.1632M} and observations \citep[e.g.][]{2014MNRAS.439..673B,2016MNRAS.455.2484N,2017A&A...602A.101R}.
In Appendix \ref{sec.Appendix.boundary}, we make detailed discussion on the influence of the variation of the resonance offset boundary.
To quantify the differences between the observational sample and randomly paired control sample, we define a metric, the normalised fraction of planet pairs being near-MMRs, $\hat{F}_{\rm MMR}$, which is mathematicalexpressed as
    \begin{equation}
        \hat{F}_{\rm MMR} =  \frac{F^{\rm obs}_{\rm MMR}}{F^{\rm con}_{\rm MMR}}.
    \end{equation}
This metric is thus a measure of how more frequently actual planets are near-MMR compared to a random paired situation.

\begin{figure}[!t]
\centering
\includegraphics[width=\linewidth]{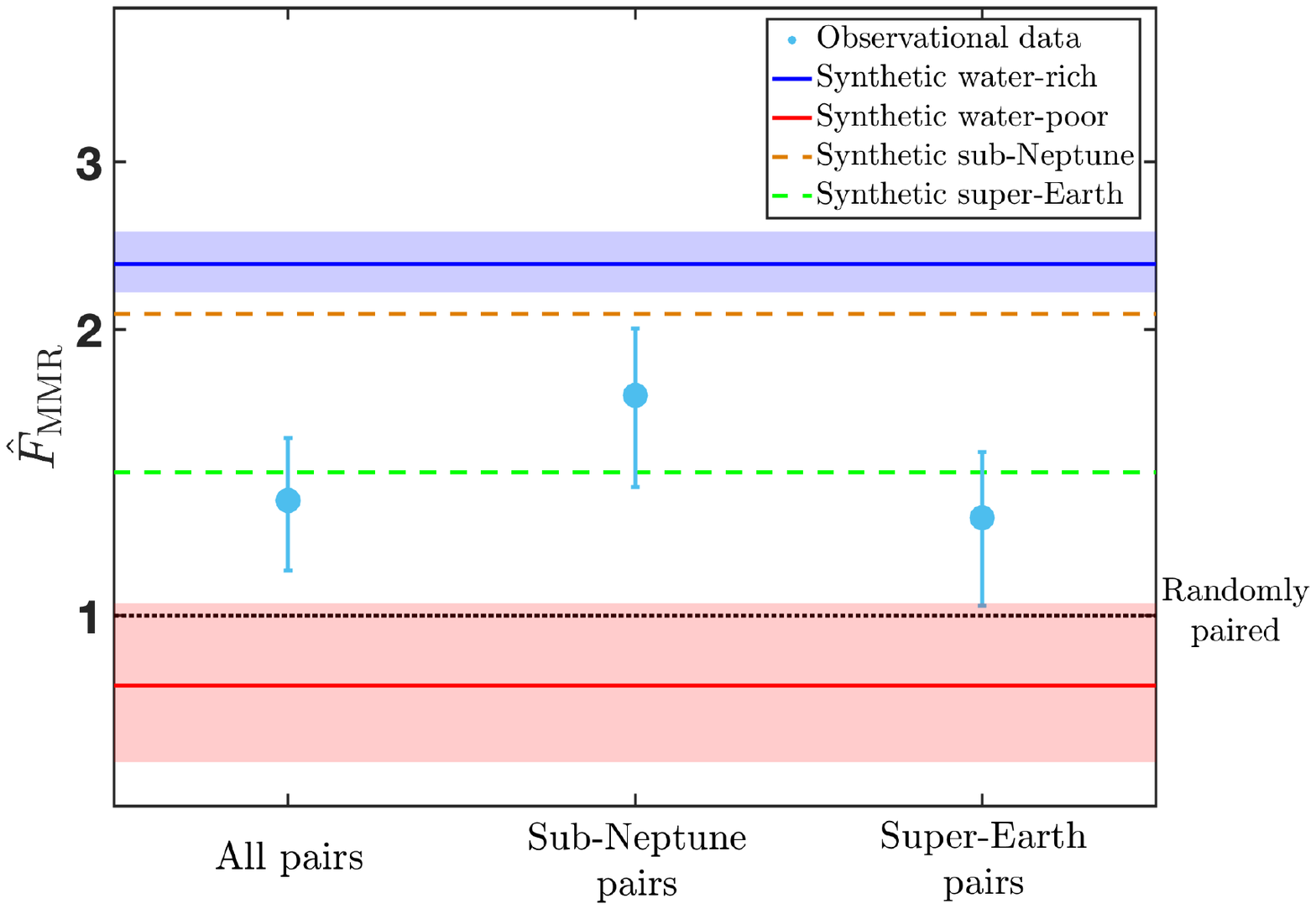}
\caption{ The normalised fraction of near-MMRs pairs, $\hat{F}_{MMR}$ (solid points) with $1-\sigma$ uncertainties (vertical bars) for all actual planet pairs, sub-Neptune pairs and super-Earth planet pairs in Kepler multiple transiting systems.
The two horizontal solid lines and shaded regions represent the theoretical predictions and $1-\sigma$ uncertainties of water-rich (blue) and water-poor planet pairs (red) derived from the synthetic sample. 
The two horizontal dashed lines denote the theoretical predictions of simulated sub-Neptune (brown) and super-Earth (green) pairs derived from the synthetic sample. 
The result of randomly paired control sample is plotted as dotted black line and by definition equal to unity.
\label{figFAMMR}}
\end{figure}

Figure \ref{figFAMMR} shows the normalised fractions in near-MMRs $\hat{F}_{\rm MMR}$ as well as their $1-\sigma$ intervals for all the actual planet pairs, sub-Neptune pairs, and Super-Earth pairs. The plot also compares to the result obtained for the synthetic population. They will be discussed in the next section.

As can be seen, as a whole, Kepler planet pairs show a statistically weak preference to be near-MMR ($\hat{F}_{\rm MMR}>1$) by a factor of { $1.3^{+0.2}_{-0.2}$ (with a confidence level of 91.21\%)} comparing to the random paired control sample, which is consistent with previous studies \citep{2011ApJS..197....8L,2014ApJ...790..146F}.
However, when splitting the actual Kepler pairs into the two sub-samples (sub-Neptunes and super-Earths), a more interesting picture arises: comparing to the control sample, sub-Neptunes pairs show a significant preference of near-MMR with a $\hat{F}_{\rm MMR}$ of { $1.7^{+0.3}_{-0.3}$}.
Out of the 10,000 sets of resampled data, $\hat{F}_{\rm MMR}$ of sub-Neptune pairs is larger than 1 for { 9,843 times, corresponding to a confidence level of 98.43\%}.
On the contrary, the super-Earth pairs have a $\hat{F}_{\rm MMR}$ of { $1.3^{+0.2}_{-0.3}$} which is consistent with that of the random paired control sample within $1-\sigma$ errorbars.
Furthermore, for the sub-Neptune pairs, there exists an obvious asymmetry in the distribution of period ratios near first-order resonances, with an excess of planet pairs lying wide of resonance by a factor of $9.5^{+13.9}_{-4.2}$ comparing to planets lying narrow of resonance with a confidence level of 99.95\%.
Whereas, for super-Earth pairs, the distribution of period ratios near first-order resonances is nearly symmetric and the number ratio of planet pairs lying wide of resonance over those lying narrow of resonance is $1.1^{+0.4}_{-0.4}$, which is similar to a random distribution.

Bases on the above analyses, we conclude that sub-Neptunes in the actual Kepler sample have a significant preference to be captured in a near mean-motion-resonance configuration.
Whereas, super-Earths are statistically randomly paired. 
This could suggest that some sub-Neptunes have experienced a different formation process than super-Earth planets where orbital migration and resonant capture played a more important role.

\section{Theoretical predictions for planet pairs for different compositions}\label{sec.PR.theory}

\begin{figure*}
\centering
\includegraphics[width=0.9\textwidth]{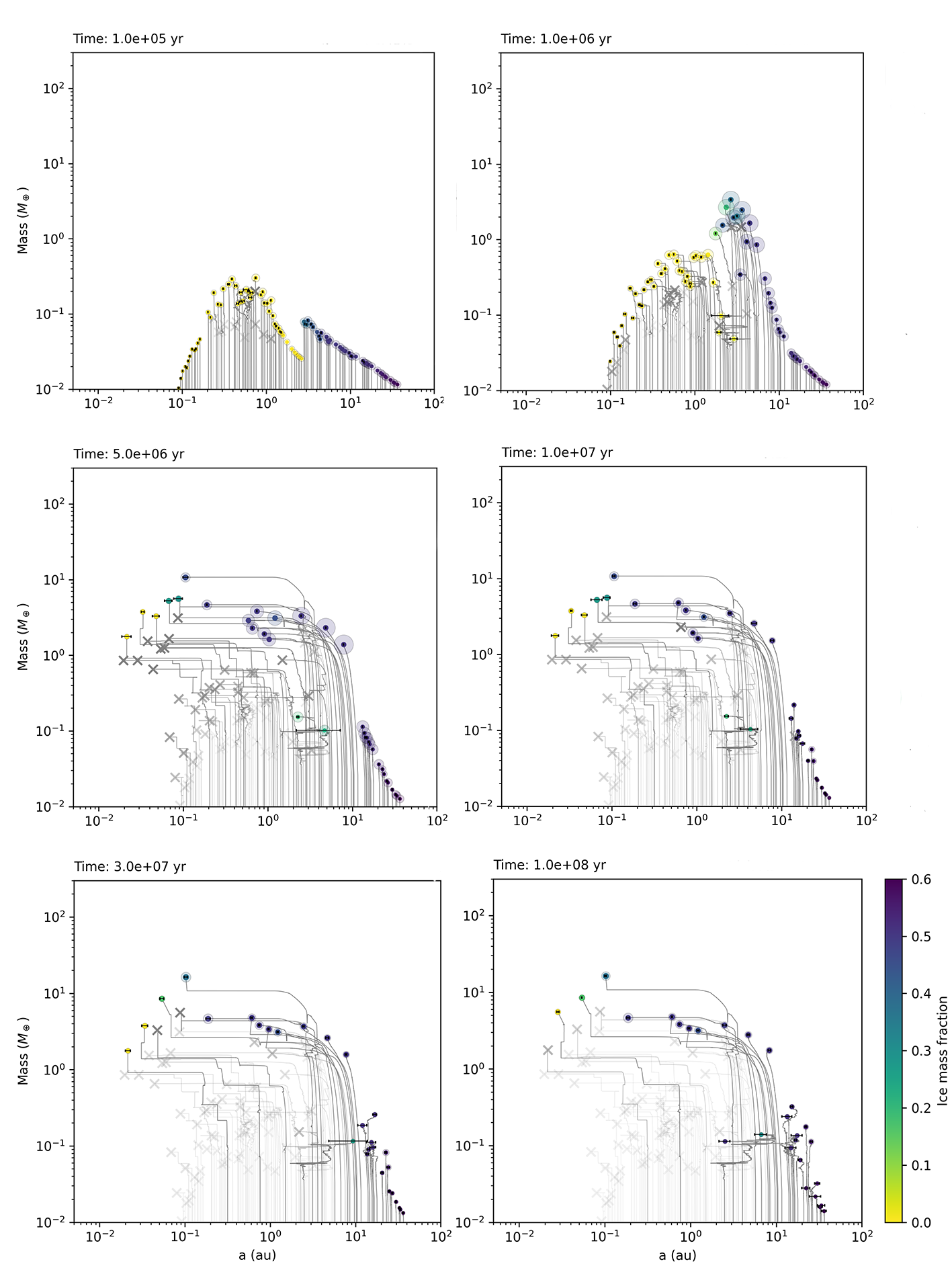}
\caption{Example of the formation of a synthetic planetary system from initially 100 lunar-mass embryos in a typical protoplanetary disk with solar metallicity ($\rm [Fe/H] = 0.0$).  The epochs in time (in years) are printed in the top-Left of the six panels. Solid points show (proto)planets with the semi-transparent part scaling with radius. Grey crosses (fading in time) show the last position of protoplanets that were accreted by other more massive bodies. The colours of points represent the mass fraction of ice in the core. Horizontal black bars go from the periastron to the apoastron, i.e. represent orbital eccentricity.  Lines show the growth tracks in the semi-major axis-mass plane. 
\label{figFEprocessBernmodel}}
\end{figure*}

In this section, we analyse the period ratio distribution of adjacent planet pairs in the synthetic population. The model provides not only the orbital parameters and radius of planets, but also their mass, composition, and dynamic and formation histories.

To illustrate how synthetic planetary systems and their system architecture emerge, we first describe the formation and evolution of one specific system out of the 1000 in the synthetic population (Fig. \ref{figFEprocessBernmodel}). The physical processes and mass scales leading to the emergence of the four different planetary system architectures were described in \cite{2023A&A...673A..78E}. For the analysis here, the Class I and Class II systems identified there are most relevant. They both contain (close-in) low-mass planets, but originating from different formation pathways.

Class I systems contain in the inner part (inside of about 1 AU) super-Earth planets that have formed approximately in situ. Thus, they have rocky (silicate-iron) cores without much water. Their formation pathway is characterized by growth initially via planetesimal accretion and then giant impacts (embryo-embryo collisions) and only limited orbital migration (only inside of the iceline). After the dissipation of the gaseous disk, atmospheric escape removes in most cases their H/He envelopes. These evaporated bare cores have radii $\le 1.7 R_\oplus$ and populate the super-Earth peak of the observed radius distribution.

Class II systems in contrast contain larger ex situ ice-rich sub-Neptunian planets that have primarily accreted beyond the iceline. They first accrete icy planetesimals outside of the water iceline at approximately constant orbital separation and then migrate inward at approximately constant mass in a ``horizontal branch'' \citep{2009A&A...501.1161M}. They contain about 50\% water in mass. Because of the runaway greenhouse radius inflation effect \citep{2020A&A...638A..41T}, they have radii $\ge 2.1 R_\oplus$ and populate the observed sub-Neptune peak. In the model, the radius gap is thus caused by the distinct formation and evolution pathways of the planets above and below it: formation (orbital migration) leading to the presence of ex situ ice-rich sub-Neptunes, and evolution (atmospheric escape) leading to the presence of approximately in situ rocky super-Earths.  

The governing mass scale in the { Class I} systems is the Goldreich mass. It corresponds to the mass resulting from the final giant impact phase where mutual scattering increase the eccentricities, which in turn defines the width of an effective feeding zone. Whereas, in the { Class II} systems it is the equality mass (when migration and planetesimal accretion timescales become equal) or the saturation mass (when the corrotation torques saturate). When these masses are reached, the planets start to migrate inwards.
In mass (instead of radius) space, the two classes overlap much more, but below (above) about 5 $M_\oplus$ super-Earth (sub-Neptunes) dominate \citep[see also][]{2020A&A...643L...1V,2020A&A...644A.174V}.


\subsection{Formation of one synthetic system}\label{sec.synthetic.system.example}
The specific example shown in Fig. \ref{figFEprocessBernmodel} displays the temporal evolution  of a planetary system in the semi-major axis-mass plane for a typical proptoplanetary disk with solar metallicity ($\rm [Fe/H] = 0.0$). The initial masses of gas and dust are 0.027 $M_\odot$ and 105 $M_\oplus$, respectively. Both these values are close to the mean values of the probability distribution of the disk initial conditions \citep{2021A&A...656A..70E}.
At $t=0$, 100 lunar mass embryos are uniformed seeded inside of 40 AU in the logarithm of the semi-major axis.

As shown in the top-left panel of Fig. \ref{figFEprocessBernmodel}, shortly after the beginning ($10^5$ yr), the dominating process is the quasi in-situ accretion of planetesimals in the feeding zone of the embryos. 
Some dynamical interactions have also started among some embryos, leading to some collisions and mergers, which are indicated by grey crosses. Orbital migration is not yet important since the timescales are quite long for very low masses, whereas planetesimal accretion is fast, inducing nearly vertical upward tracks in the semi-major axis-mass plane. The specific pattern of the maximum mass as a function of distance can be understood in the following way: from 0.1 AU to the largest masses at 0.4-0.8 AU, the masses are increasing, because planets are (except for giant impacts) stuck at the local planetesimal isolation mass which is a increasing function of distance for the minimum-mass solar nebula (MMSN)-like planetesimal surface profile assumed here \citep{1993ARA&A..31..129L}.
Outside of this, the masses are first decreasing again. This is the consequence of the increase of the growth timescale with orbital distance for oligarchic growth \citep{2003Icar..161..431T}. Slightly inside of 3 AU, there is again a sudden increase in the masses. This is caused by the water iceline which increases the planetesimal surface density approximately by a factor 2. Even further out, the masses decrease, which is again caused by the slower planetesimal accretion at larger distances.

At 1 Myr (the top-right panel), (just) beyond the ice line, a handful protoplanets with masses between about 1 and 4 $M_\oplus$ have formed and some inward migration has started, leading to the tracks bending inward. This occurs at the saturation mass where the corrotation torque saturates ($M_{\rm sat}$ or at the equality mass where the migration timescale becomes shorter than growth timescale \citep[see][]{2023A&A...673A..78E}. For protoplanets inside the ice line, the growth via giant impacts (embryo-embryo collisions) now plays a more important role and has allowed some protoplanets in the inner disk to grow beyond the local isolation mass. We also see that some orbital migration has started also here.

At 5 Myr (the middle-left panel), in the outer disk, the massive protoplanets initially formed beyond the ice lines have reached masses of several Earth masses and migrated further inward. In this configuration (Type I migration), outer more massive planets generally migrate faster and the protoplanets capture each other in resonant convoys and migrate together \citep[e.g.][]{2008A&A...483..633P,2013A&A...558A.109A}. During their migration inward, the icy planets keep almost constant masses, leading to tracks in the ``horizontal branch'' \citep{2009A&A...501.1161M}: this is due to the fact that the planetesimals inside of their position were already accreted by the embryos growing there. The outer ice planets have pushed three inner rocky protoplanets to the inner edge of the disk at about 0.02 AU. In this inner disk, the growth via giant impacts becomes the dominant effect among the rocky planets. In particular, tens of protoplanets with masses of a few percent to tenths of $M_\oplus$ that existed at 1 Myr have now formed these  three super-Earths via numerous giant impacts.

After the disk disappears ({\cdc $\sim 5.6$ Myr}), we continue to make N-body dynamic interaction to 100 Myr. 
As can be seen in the last three panels, during 10-100 Myr, the dynamic interaction between (proto)planets have induce further giant impacts and scatterings, reducing the number of planets and destroying mean-motion resonance (MMR), especially for the super-Earths at very close orbits.

This system illustrates how in the synthetic population the larger sub-Neptune planets initially formed beyond the ice-line with large mass fractions of water/ice in their interiors are expected to have a larger fraction of MMRs compared to the smaller volatile-poor rocky Super-Earth formed inside the ice-line.

While the formation pathway dominated by inward migration characterising this system corresponds to the Class II, in this specific system, one inner rocky planet has remained. This is in contrast to most Class II systems, where only ice-rich sub-Neptunes remain. The compositional ordering with rocky planets inside and ice ones outside is rather characteristic of Class I systems with little orbital migration. Thus, this system has partially a mixed character. It is an interesting example with planets spanning the valley. A number of such systems are observationally known \citep{2012Sci...337..556C,2020MNRAS.491.5287O}. 
{\cdc Furthermore, this system contains seven planets (five sub-Neptunes and two super-Earths) with periods between 1 and 400 days.
After applying apply the synthetic detection bias using the KOBE program, three of them (two sub-Neptunes and one super-Earth) can be observed.
Interestingly, the two sub-Neptunes are both born beyond the ice-line and contains significant amount (more than 40\%) of water in their envelopes (Class II), while the super-Earth are born at $\sim 0.8$ AU (inside the ice-line) and the mass fraction of water in its envelope is only $\sim 1\%$. 
That is to say, water-rich planets and water-poor planets locate on the two sides of the radius valley.}

\subsection{Statistical analysis of the synthetic systems}
We next study whether the different formation pathways of planets in the synthetic system are reflected in the frequency of pairs in MMRs.

\begin{figure}[!t]
\centering
\includegraphics[width=1.1\linewidth]{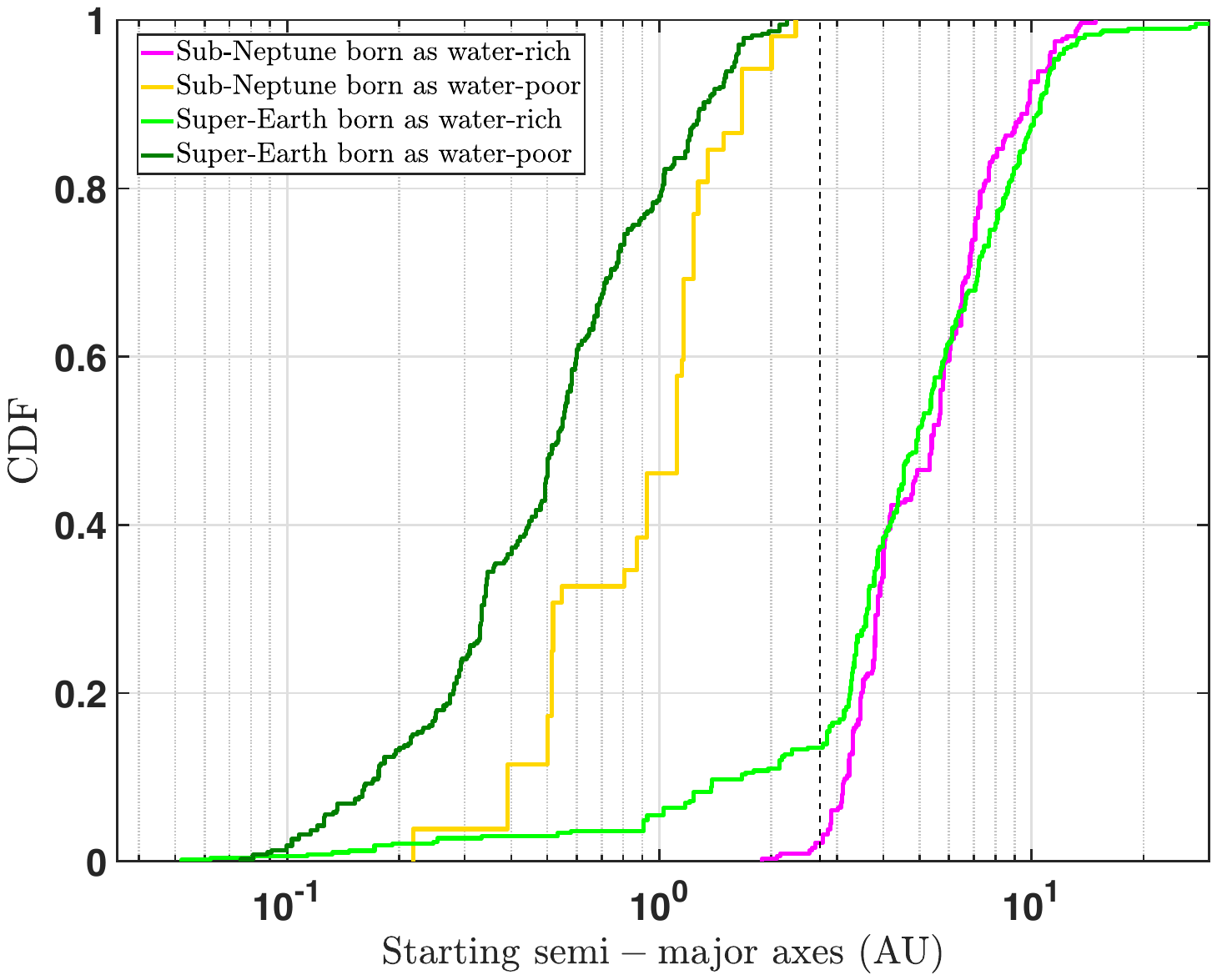}
\caption{The cumulative distributions of the starting position of the embryos  that eventually became planets with different compositions and radii.
The dashed line represents the typical snow line location.
\label{figastartplanets}}
\end{figure}

We evaluate the $Z$ value at the moment when the parent protoplanetary disk disappears. We divide the planets into two different categories: First, $Z \ge 0.1$ water-rich planets\footnote{When the envelope mass is small compared to the total mass of the planet (or even zero in case that the envelope was fully evaporated during the long-term evolution), then the absolute amount of ice these planets contain is low or even zero at 5 Gyr. However, we still refer to these planets for simplicity as ``ice-rich'' understanding this as a sign that orbital migration was important.} corresponding to bodies that have accreted ice-rich material, either directly by accreting water-rich planetesimals or indirectly by colliding with a water-rich protoplanet. The ice content serves here as a proxy for formation pathways where orbital migration was of import (Class II). Second, water-poor ($Z < 0.1)$ planets that have mainly sourced the region inside of the iceline, and for which migration was less important (Class I architectures). 

Figure \ref{figastartplanets} displays the cumulative distributions of the starting positions for planets of different radii and compositions.
As can be seen, as expected, water-rich planets are mainly formed beyond the ice-line, while water-poor planets are  formed inside the ice-line.
In the selected synthetic sample, sub-Neptune are in their large majority water-rich (706 of 740, i.e. 95\%).
Super-Earths consists of 437 water-rich planets and 483 water-poor planets. The former are planets that originally had an envelope with a $Z \ge 0.1$ that was subsequently lost.
{\cdc The selected synthetic sample contains { 796 stable adjacent planet pairs, which are composed of 596 pairs of two water-rich planets (hereafter as water-rich pairs), 135 pairs of two water-poor planets (hereafter as water-poor pairs) and 65 pairs of one water-rich planet and one water-poor planet (hereafter as mixed pairs).
That is to say, most (over 90\%) of planets are adjacent to planets with the same composition category.}}


\begin{figure}[!t]
\centering
\includegraphics[width=
\linewidth]{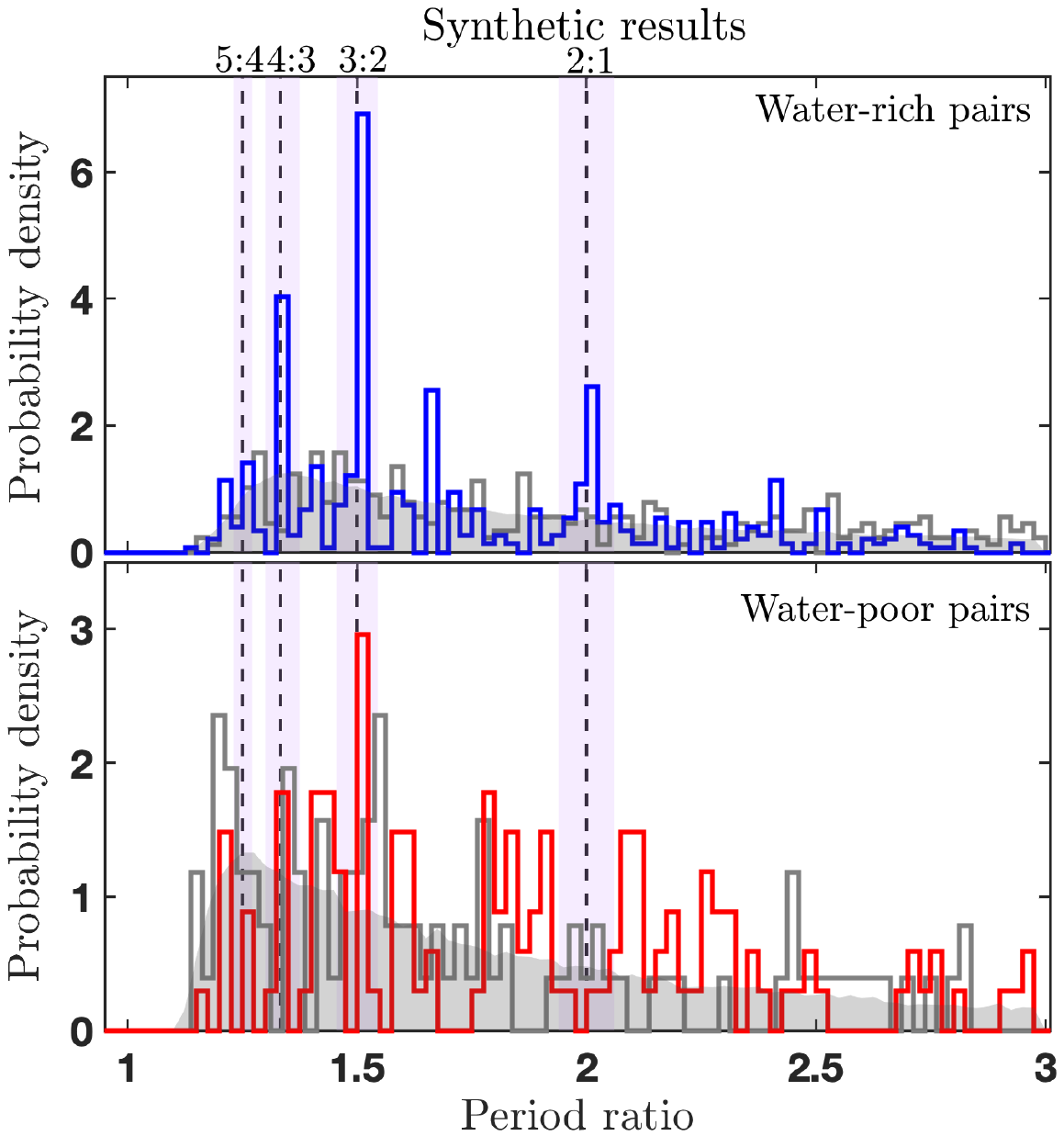}
\caption{{Probability density functions of period ratios of adjacent water-rich (top panel) and water-poor (bottom panel) planet pairs derived from the synthetic sample and the corresponding control sample with the assumption that planets are randomly paired (grey region for the control sample by combining the 1,000 times of randomly pairing and grey line is one example with similar fractions in near-first-order MMRs in the 1,000 sets).} 
\label{figPRhistModel27}}
\end{figure}

We then investigate the period ratios for adjacent planet pairs in the synthetic sample with the same method as described in Sect. \ref{sec.PR.obs} for the actual observed planets.
Figure \ref{figPRhistModel27} compare the probability density distributions of period ratios derived from the synthetic sample and the control sample for planet pairs of different compositions.
To quantify the preference in near-MMRs, we also calculate their normalised fractions in near-MMR, $\hat{F}_{\rm MMR}$ relative to the random paired sample.
As shown in Fig. \ref{figFAMMR} by the blue line and region, water-rich planet pairs exhibit a significant preference to be near-MMR compared to the randomly paired control sample by a factor of { $2.3^{+0.2}_{-0.2}$ with a confidence level $\gtrsim 5-\sigma$}.
On the contrary, water-poor planet pairs (red line and region) have a normalised fraction in near-MMRs ({ $0.8^{+0.2}_{-0.1}$}) that a bit lower than (but statistically indistinguishable) that of random paired control sample.

Since the presence of water cannot be directly observed, we also calculate $\hat{F}_{\rm MMR}$ of some subgroups distinguishing (synthetic) super-Earths and sub-Neptunes. The derived results are { $2.2^{+0.3}_{-0.2}$, $2.1^{+0.3}_{-0.3}$, $0.8^{+0.6}_{-0.4}$, and $0.7^{+0.2}_{-0.2}$} for water-rich sub-Neptune pairs, water-rich super-Earth planet pairs, water-poor sub-Neptune pairs and water-poor Super-Earth pairs respectively. 
These values are consistent with the results from the whole water-rich and water-poor planet pairs within $1-\sigma$ uncertainties.

{\cdc To directly compare to the observational results, we also divide the synthetic sample into sub-Neptunes and super-Earths only according to radius, { yielding 154 adjacent sub-Neptune pairs (146 water-rich, 6 water-poor, 2 mixed) and 273 super-Earth pairs (121 water-rich, 137 water-poor, 15 mixed)}.
Then we calculate their $\hat{F}_{\rm MMR}$, which are { $2.1^{+0.2}_{-0.2}$ and $1.4^{+0.2}_{-0.1}$}, respectively.
As shown in Fig. \ref{figFAMMR}, the synthetic results for sub-Neptune (brown) and super-Earth (green) pairs are higher than the observational results, which imply that the planet population generated by the Bern model contains larger fractions of water-rich planets comparing to the Kepler sample (see Sect. \ref{sec.dis.implication} for detailed discussions).}


\section{Implication on the formation and composition of planets}
\label{sec.dis.implication}
In this subsection, we compare the observational results (Sect. \ref{sec.PR.obs}) with the theoretical predictions (Sect. \ref{sec.PR.theory}) of the period ratio distributions.
From the comparison of the model and observed $\hat{F}_{\rm MMR}$, we can also put a nominal prediction on the proportion that formed as water-rich (or more exactly ex-situ) planets $f_{\rm rich}$ with the following formula:
\begin{equation}
    f_{\rm rich} = \frac{\hat{F}_{\rm MMR}-\hat{F}^{\rm poor}_{\rm MMR}}{\hat{F}^{\rm rich}_{\rm MMR}-\hat{F}^{\rm poor}_{\rm MMR}},
\end{equation}
{ where the `rich' and `poor' superscripts denote results derived respectively from water-rich and water-poor planets in the synthetic sample.}
{ The uncertainty of $f_{\rm rich}$ is calculated from the uncertainties of ${\hat{F}_{\rm MMR}}$, $\hat{F}^{\rm rich}_{\rm MMR}$ and $\hat{F}^{\rm poor}_{\rm MMR}$ by means of error propagation.
Here we ignore the contributions of mixed pairs as these are rare ($\lesssim 5\%$) in simulated samples.
To verify the Eq. (7), we calculate $f_{\rm rich}$ for the the sub-Neptune and super-Earth pairs in the synthetic sample generated by the Bern models from their period ratio distributions.
{ The resulting $f_{\rm rich}$ are $87^{+13}_{-20}\%$ and $40^{+22}_{-17}\%$, which are very similar to the fractions of sub-Neptunes ($706/740 =95\%$) and super-Earths ($437/920 = 48\%$) as water-rich in the synthetic sample.}}

The derived constraints on the formation and composition of Kepler-planets are as follows:

\begin{enumerate}

\item For the sub-Neptune pairs, the actual Kepler sample shows that they exhibit a preference to be near-MMR. Quantitatively, the derived observational { $\hat{F}_{\rm MMR}$, $1.7^{+0.3}_{-0.3}$}  is smaller than the $\hat{F}_{\rm MMR}$ derived for synthetic ex-situ water-rich planet pairs, { $2.3^{+0.2}_{-0.2}$} but larger than synthetic in-situ water-poor planet pairs, { $0.8^{+0.2}_{-0.1}$}.
In the 10,000 sets of resampled data, the derived $\hat{F}_{\rm MMR}$ from the observational sample is smaller than those of water-rich planet pairs for { 9,460 times, and lager than water-poor planet pairs for 9,878 times, corresponding to confidence levels of 94.60\% and $98.78\%$, respectively}.
That is to say, the hypothesis that observed sub-Neptunes are all water-rich or water-poor could be confidently ($\gtrsim 2 \sigma$) rejected.

We also calculate the $f_{\rm rich}$ using Eq. (7) and the resulting $f_{\rm rich}$ is { $60^{+20}_{-28}\%$} for sub-Neptunes in Kepler multiple transiting systems. Taken at face value, this would mean that formation-wise the actual sub-Neptunes are approximately a half-half combination of two populations that formed about in an in/ex situ way. \\

\item For the Super-Earth pairs in actual Kepler multiple systems, we find that they are statistically randomly paired and show no preference to be captured in near-MMR.
Quantitatively, $\hat{F}_{\rm MMR}$ derived from the Kepler data, { $1.3^{+0.2}_{-0.3}$, is a bit higher than the theoretical prediction of water-poor planet pairs, $0.80^{+0.2}_{-0.1}$ but significantly smaller than that of water-rich synthetic super-Earth planet pairs, $2.3^{+0.2}_{-0.2}$ with a confidence level of 99.82\%.}
The derived $f_{\rm rich}$ from Eq. (7) is { $28^{+21}_{-17}\%$} for Super-Earth pairs, demonstrating that the observed super-Earth planets are mainly ($\sim 50\%-90\%$) born as water-poor (i.e. iron plus silicate) cores inside the ice-line. \\

\cite{2020ApJ...893...43M} and \cite{2023AJ....165..174W} propose a hypothesis which could also contribute to the lower fraction in near-MMR for Super-Earths.
If a planetary system formed around a star with strong XUV emissions, planets born with significant gas envelope may lose their envelope via photoevaporation and evolve to bare super-Earth planets \citep[e.g.][]{2013ApJ...775..105O,2014ApJ...795...65J}.
Their semi-major axes could be impulsively changed if they quickly lost $\gtrsim 10\%$ of their total mass within a timescale $\lesssim 10^4-10^5$ years.
In this way, some pairs may escape from MMR configurations, further reducing the fraction of super-Earths near MMRs. Therefore, considering the effect of mass-loss on the planetary orbits\footnote{Our model currently does not include evaporation and N-body interactions at the same time, therefore it does not include the effect that atmospheric escape may break MMRs}, the theoretical expectation of water-poor super-Earth pairs in near-MMRs should be lower.
According to Eq. (7), the fraction of super-Earth planets that formed as water-rich in Kepler transiting multiple systems would be higher, thus the above $f_{\rm rich}$ ($28^{+21}_{-17} \%$) is somewhat a lower limit. \\

\end{enumerate}

In Appendix \ref{sec.Appendix.boundary}, we also calculate the $\hat{F}_{\rm MMR}$ and $f_{\rm rich}$ for the Kepler by taking different boundaries of the resonance offset as near-MMR.   
The derived results (Fig. \ref{figFAMMR_Deltaboundaries}) are well consistent with the above results (Fig. \ref{figFAMMR}), demonstrating that the variation of boundary of resonance offset have little influence on our main conclusions.





\section{Summary}
\label{sec.Summary}
Based on the Kepler DR 25 catalogue \citep{2017ApJS..229...30M}, we have investigated the distributions of orbital period ratios of adjacent planet pairs in the multiple transiting Kepler systems, separating the sample into super-Earth and sub-Neptune pairs.
We find that sub-Neptune pairs shows a significant preference ({ 98.43\%}) to be captured in near-MMRs configurations by a factor of { $1.7^{+0.3}_{-0.3}$} compared to a random paired control sample, while the super-Earth planet pairs show no significant difference from a random distribution (Sect. 3, Fig. \ref{figPRhistobservation}).

Then, using a synthetic planetary population generated by the Generation III Bern model \citep{2021A&A...656A..69E,2021A&A...656A..70E}, we study the frequency of MMRs in the two classes of close-in planets identified in \cite{2023A&A...673A..78E} that differ by their formation pathways: Class I approximately in-situ water-poor planets that have formed within the ice-line and Class II ex-situ water-rich planets that were born beyond the ice line and have migrate inward (see Sect.  4 and Fig. \ref{figFEprocessBernmodel}-\ref{figastartplanets}). For Class I a final giant impact phase is governing the resulting architecture (orbits and masses) whereas for Class II the effects of Type I orbital migration is dominant. The two classes correspond mainly to super-Earth (Class I) and sub-Neptunes (Class II). As for the actual Kepler planets, we derive their period ratio distributions and normalised fractions that are near-MMRs (Fig. \ref{figPRhistModel27}). The prevalence of near-MMRs relative to the random paired sample are { $2.3^{+0.2}_{-0.2}$ and $0.8^{+0.2}_{-0.1}$} for the synthetic sub-Neptunes and super-Earths, respectively.

By comparing the observational results with the theoretical predictions (Sect. 5 and Fig. \ref{figFAMMR}), we reject the hypothesis that the actual sub-Neptunes in Kepler multiple transiting systems are only water-rich ex-situ or only water-poor in-situ planets with a confidence level of $\sim 2-\sigma$. Instead, our result suggest that they should be made from a mixture of both the two populations/two formation pathways. Our nominal estimation is that { $60^{+20}_{-28}$\%} of the actual sub-Neptunes have formed via a formation pathways where orbital migration was important, potentially implying an origin beyond the iceline and thus an ice-rich composition.  

In contrast, actual super-Earth planet pairs have a normalised fraction of near-MMR that is significantly ({ 99.82\%}) smaller than that of synthetic water-rich planet pairs but statistically consistent with that of synthetic water-poor planet pairs within $1-2 \sigma$ errorbars. This suggests that the majority of actual super-Earth planets correspond to water-poor planets born within the ice-line with no/limited importance of orbital migration.

Our results on observed MMRs thus suggest that  close-in sub-Neptunes and super-Earth had partially different formation pathways. This supports a view \citep{2024NatAs.tmp...33B} where the radius valley is both a consequence of both formation \citep[orbital migration; larger water-rich sub-Neptunes born ex-situ versus smaller rocky super-Earth born in-situ; see also][]{2020A&A...643L...1V,2020A&A...644A.174V} as well as evolution (atmospheric escape) that populates the super-Earth peak in the radius distribution.  

Future studies, both from observational analyses \citep[e.g. larger sample of planets in multiple systems, studies of atmospheric composition;][]{2006SSRv..123..485G,2014ExA....38..249R}
and from numerical simulations of theoretical formation and evolution models, will test our results and further provide more clues on the formation, evolution, and compositions of close-in low-mass exoplanets, the most frequent currently known type of planets.

\section*{Acknowledgments}
This work is supported by the National Key R\&D Program of China (No. 2019YFA0405100, 2019YFA0706601) and the National Natural Science Foundation of China (NSFC; grant No. 11933001, 12273011).
We also acknowledge the science research grants from the China Manned Space Project with NO.CMS-CSST-2021-B12 and CMS-CSST-2021-B09. 
J.-W.X. also acknowledges the support from the National Youth Talent Support Program.
D.-C.C. also acknowledges the Cultivation project for LAMOST Scientific Payoff, Research Achievement of CAMS-CAS and the fellowship of Chinese postdoctoral science foundation (2022M711566). This research was supported by the Excellence Cluster ORIGINS which is funded by the Deutsche Forschungsgemeinschaft (DFG, German Research Foundation) under Germany's Excelence Strategy – EXC-2094-390783311.
C.M. acknowledges the support from the Swiss National Science Foundation under grant 200021\_204847 ``PlanetsInTime''. 
Parts of this work has been carried out within the framework of the NCCR PlanetS supported by the Swiss National Science Foundation under grants 51NF40\_182901 and 51NF40\_205606.

\appendix
\renewcommand\thefigure{\Alph{section}\arabic{figure}}
\setcounter{figure}{0} 

{

\section{Influence of the limit for Hill stability}
\label{sec.Appendix.Hillstable}
In the main text, we adopt the Hill-stable criterion to ensure the generated planet pairs in the control sample is stable.
However, the expected value for the limit of Hill-stability $K$ is still uncertain.
Specifically, the minimum separation required to be Hill stable is $2\sqrt{3}$ when considering two planets in circular and coplanar orbits \citep{1993Icar..106..247G}.
\cite{2015ApJ...807...44P} find that $K \sim 10-12$ is required for
long-term stability for the adjacent planets in Kepler systems showing four or more transiting planets.
\cite{2019MNRAS.490.4575H}  generated sets of simulated `Kepler-like' planetary systems and reported a best fit of $K \sim 8$ comparing to the actual observational data.
\cite{2024AJ....167...46D} carried out numerical simulations and provided a median value of $K = 7.17$ from simulated planet pairs in long-stable planetary systems.
Thus, generally, the previous estimates for $K$ are in the range of $\sim 2\sqrt{3}-12$.

To evaluate the influence of the variation of the limit for Hill-stability for the control sample, we re-select planet pairs as Hill-stable ($H>K$ \& $H_{\rm in}+H_{\rm out}>18$) in the observed and synthetic samples with different values of $K$.
Figure \ref{figCDFOC_Ks} shows the cumulative distributions of the period ratio of planet pairs  selected as Hill stable with five typical values of $K$. 
We also perform the two-sample Kolmogorov–Smirnov (K-S) test between the distributions of period ratio derived from the observed and control samples.
We note that the sample size of the control sample is much larger than the observed pairs, which could affect the resulting $p-$ values. Thus, to further eliminate the effect of the difference in sample size, we select one set with similar fractions in the near-MMRs from the 1,000 sets in the randomly pairing procedure of the control sample (which has a same size as the observed sample before Hill-stable test) and then make the KS test.

Figure \ref{figPvalue_K} displays the resulted $p-$values as a function of $K$.
As can be seen, when $K$ is small ($\lesssim 5$), the control sample has more planet pairs with period ratio less than 1.5. 
Whereas, when K exceeds $\sim 10$, most of planet pairs with period ratios $< 1.5$ becomes `unstable' in the observed sample, indicating that the theoretical limit of $\gtrsim 10$ is too strict for our selected Kepler sample.
Besides, the control sample generally have a larger distribution in period ratio comparing to the observed sample. 
Thus, the resulted $p-$values first increase and then decrease.
Thus, we take K with maximum p-value, 7.1 as the best value for the control sample.
This result is consistent with previous estimates in the range of $\sim 5-8$ \citep{2015ApJ...808...71M,2019MNRAS.490.4575H,2024AJ....167...46D} but is smaller than that of \cite{2015ApJ...807...44P} ($\sim 10-12$). 
This is not expected because our selected observed sample mainly consists of systems with two or three transiting planets and the theoretical limit $K$ would be smaller than that of \cite{2015ApJ...807...44P} due to the decrease of planet multiplicity.

\begin{figure*}
\centering
\includegraphics[width=\textwidth]{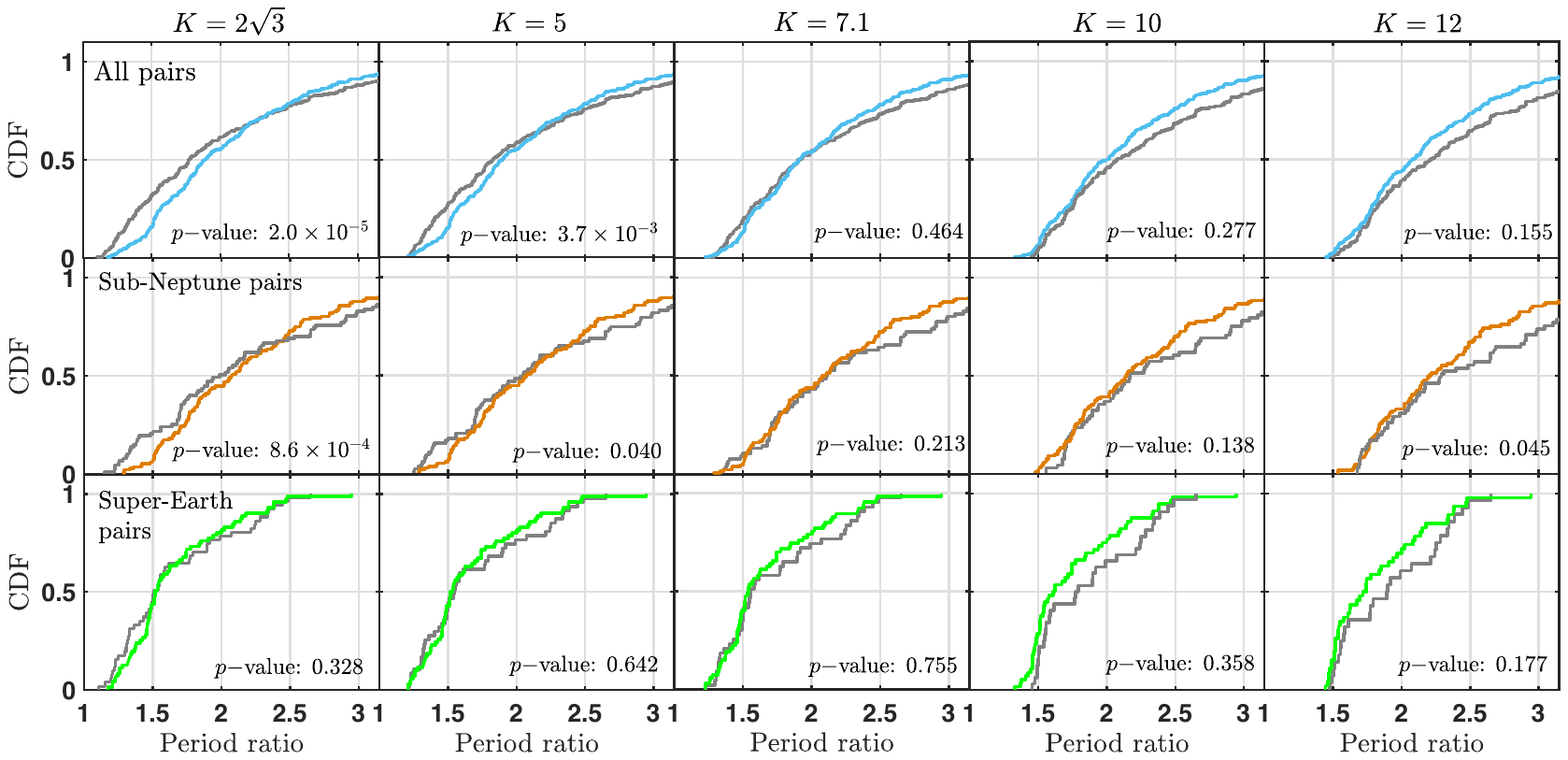}
\caption{ The cumulative distributions for the all the adjacent planet pairs (Top, cyan), sub-Neptune pairs (Middle, orange) and Super-Earth pairs (Bottom, green) selected as Hill-stability with different lower limit of orbital spacing in the observed sample. The distributions of corresponding randomly paired control samples are plotted in grey colour. We also print all the KS p-values of the observed planet pairs comparing to those in the control samples.
\label{figCDFOC_Ks}}
\end{figure*}

\begin{figure}[!t]
\centering
\includegraphics[width=\linewidth]{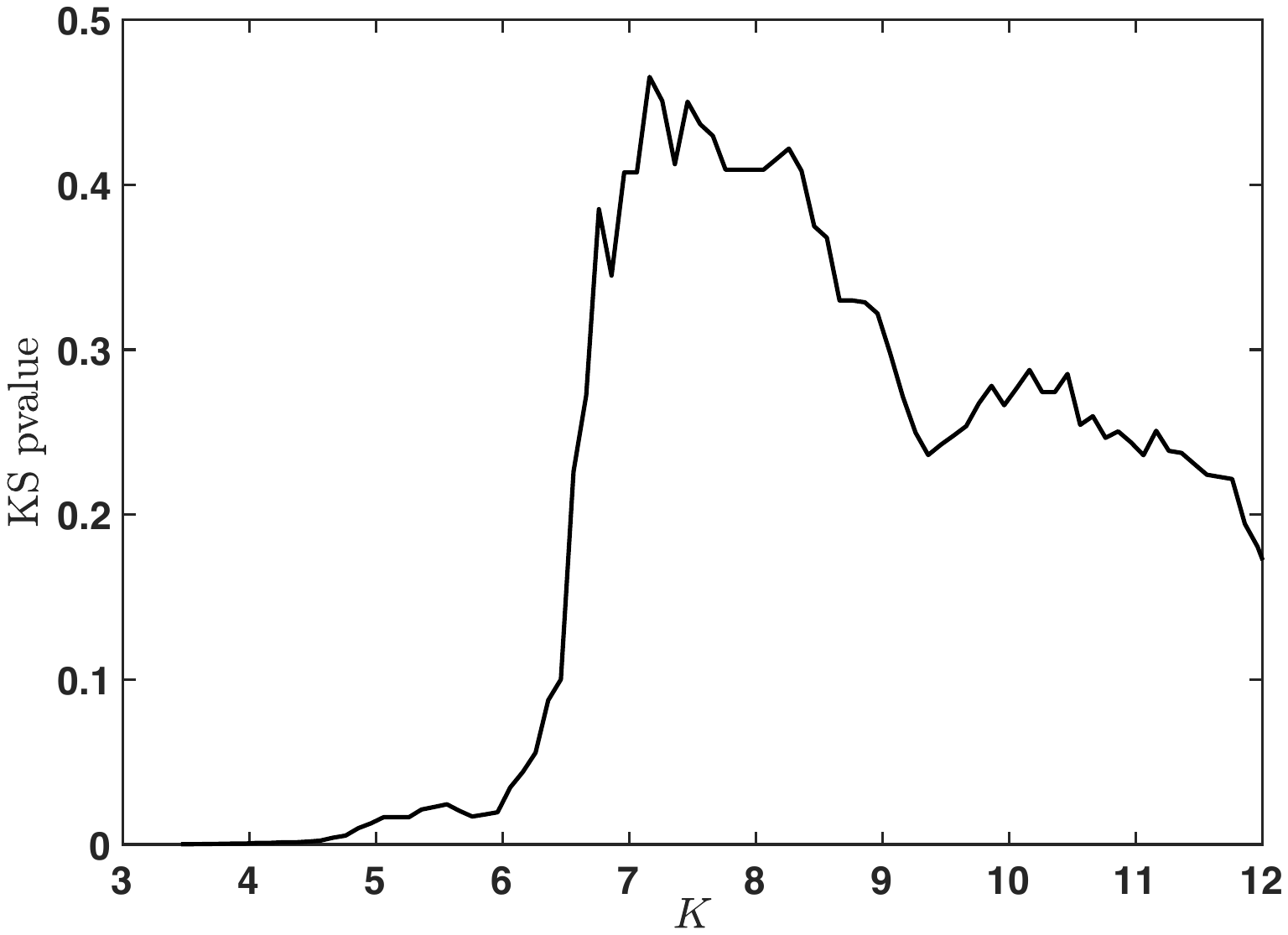}
\caption{ The KS p-values between the distributions of the period ratio of all the adjacent planet pairs for the observed and control sample when taking different limit Hill-stability $K$.
\label{figPvalue_K}}
\end{figure}

To further quantify the influence of the variation of $K$, we re-select planet pairs with $H>$ $2\sqrt{3}$, 5, 10 and 12 as Hill-stable from the observed and control samples. 
For systems with three or more planets, we also require $H_{\rm in}+H_{\rm out}>18$.
Then we calculate the normalised fraction in MMRs for the observed sample and comparing with those of water-rich and water-poor pairs in the synthetic sample.

Figure \ref{figFAMMR_K} shows the normalised fraction in MMRs for the observed and synthetic samples of different $K$.
As can be seen, all the cases show similar results:  
the sub-Neptune pairs in Kepler multiple systems have a $\hat{F}_{\rm MMR}$ smaller than that of synthetic water-rich planet pairs but larger than that of the synthetic water-poor planet pairs, while for the Super-Earth pairs, Kepler data shows a $\hat{F}_{\rm MMR}$ that statistically indistinguishable from that of the synthetic water-poor planet pairs but smaller than that of the synthetic water-rich planet pairs.
Furthermore, the proportions of sub-Neptunes and Super-Earths to be water rich are $(48^{+24}_{-26}\%, \ 27^{+12}_{-22}\%)$, $(65^{+20}_{-32}\%, \ 30^{+20}_{-25}\%)$, 
$(63^{+23}_{-43}\%, \ 23^{+28}_{-20}\%)$, and
$(64^{+25}_{-46}\%, \ 29^{+32}_{26}\%)$,
when $K$ is set as $2\sqrt{3}$, 5, 10, and 12 respectively, which are all consistent with the results of $K$ as 7.1 within $1-\sigma$ uncertainties.

\begin{figure}[!t]
\centering
\includegraphics[width=\linewidth]{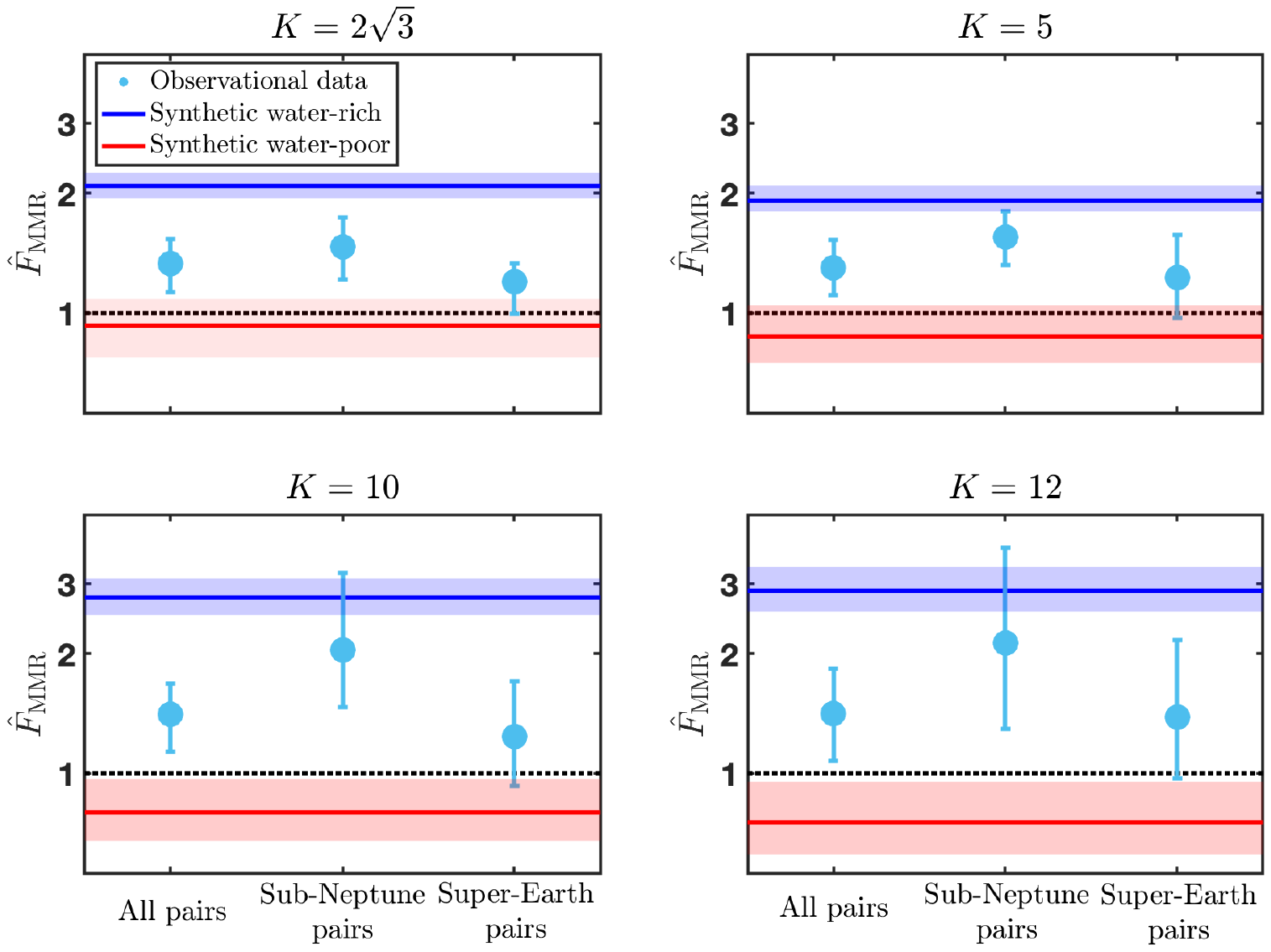}
\caption{ Similar to Fig. \ref{figFAMMR} in the main text but with different limit of Hill-stability $K$.
\label{figFAMMR_K}}
\end{figure}

Based on the above analyses, we conclude that the variation of the limit of Hill-stability has no significant effect on our main conclusions (Fig. \ref{figFAMMR}, Sect. 5).
}

\section{Influence of the boundaries of the resonance offset as near-MMR}
\label{sec.Appendix.boundary}
In this paper, we adopt the resonance offset $\Delta \equiv \frac{j}{j+1} {\rm PR}-1$ to indicates the distance of a planet pair with a given period ratio (PR) from a $j:k$ resonance.
In the aforementioned investigation in Sect. 3-4, we set 0.03 as the boundary of $\Delta$ to select those near-MMR, following previous studies \citep[e.g.][]{2014ApJ...790..146F,2020AJ....160..180J} that select 0.03 as the boundary since the overabundance of planet pairs just outside  resonances are mainly within this interval.

However, the expected value of the boundary depends on the planetary masses and eccentricities and is not certain.
From theory, N-body simulations of resonance captures generally provide a prediction of $\Delta \sim 10^{-1}$ \citep[e.g.][]{2015MNRAS.453.4089S} and the offset could further increase when considering the effect of more complex disk models on migration of planetary pairs \citep[e.g, photo-evaporation, opacity laws;][]{2015MNRAS.453.1632M,2016MNRAS.458.2051M} or some dissipative evolutions, such as tidal dissipation \citep{2012ApJ...756L..11L,2013ApJ...770...24P,2014ApJ...786..153X}. 
From observations it is found that  $\Delta \sim$ several of $10^{-2}$ \citep[e.g.][]{2014MNRAS.439..673B,2016MNRAS.455.2484N}.
\cite{2017A&A...602A.101R} investigated the Kepler systems and suggested a possible smooth trend in which $\Delta$ decreases with $P$ and the boundaries of $\Delta \sim 0.02-0.04$ for $P$ in range of 5 and 400 days (see Fig.1 of their paper).

To evaluate the effect of the variation of the resonance offset boundary on our results, we re-select planet pairs with $\Delta <$ 0.02 and 0.04 and period ratio within the closest third-order resonances as near-MMRs.
With the above new criteria, by adopting the same procedure as in Sect. 3 and 4, we recalculated the normalised fractions in MMRs for the observational sample and compare with the theoretical predictions derived from the synthetic sample.

\begin{figure}[!t]
\centering
\includegraphics[width=0.9\linewidth]{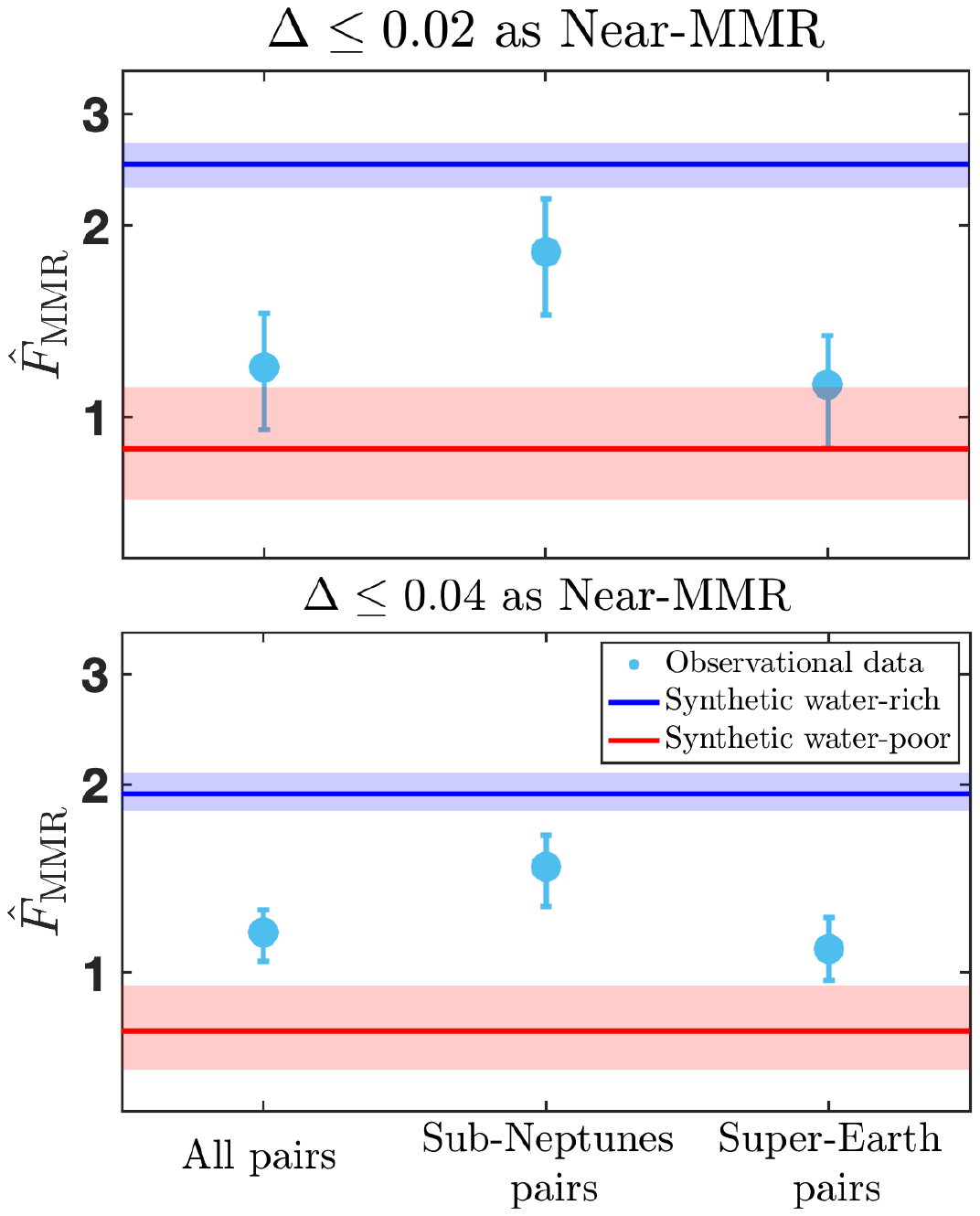}
\caption{ Similar to Fig. \ref{figFAMMR} in the main text but the boundaries of resonance offset $\Delta$ are set as 0.02 (top panel) and 0.04 (bottom panel).
\label{figFAMMR_Deltaboundaries}}
\end{figure}

Figure \ref{figFAMMR_Deltaboundaries} displays the results when the boundaries of $\Delta$ are set as 0.02 (top panel) and 0.04 (bottom panel).
As can be seen, the derived $\hat{F}_{\rm MMR}$ generally decreases with increasing $\Delta$ boundary for the observed sub-Neptune pairs and synthetic water-rich planet pairs, but change little for the observed Super-Earth pairs and synthetic water-poor planet pairs.
This result is as expected because sub-Neptune (water-rich planet) pairs in the observational (synthetic) sample exhibit overabundance just outside the exact resonances, which however could not be seen in the random paired control sample (see Fig. \ref{figPRhistobservation}, \ref{figPRhistModel27}).
Thus, when the $\Delta$ boundary expand, a larger proportion of planet pairs in control samples will be selected as near-MMR comparing to the sub-Neptunes (water-rich planets) in the observational (synthetic) sample, resulting in a smaller $\hat{F}_{\rm MMR}$.
Nevertheless, the main conclusions of our work (see Sect. 5) maintain.
{ Specifically, for the sub-Neptune pairs in Kepler multiple systems, when the boundary is set as 0.02 (0.04), the derived $\hat{F}_{\rm MMR}$ is smaller than that of synthetic water-rich planet pairs with a confidence level of 92.15\% (95.77\%), but larger than that of the synthetic water-poor planet pairs with a confidence level of 97.24\% (98.37\%).}
While for the Super-Earth pairs, Kepler data shows a $\hat{F}_{\rm MMR}$ that statistically indistinguishable ($\lesssim 1-\sigma$) from that of the synthetic water-poor planet pairs but significantly smaller than that of the synthetic water-rich planet pairs with a confidence level of 99.91\% (99.22\%) when boundary is set as 0.02 (0.04).
Furthermore, the proportions of sub-Neptunes and Super-Earths to be water rich { are $(54^{+25}_{-27}\%, \ 7^{+23}_{-7}\%)$ and $(60^{+23}_{-24}\%, \ 25^{+18}_{-12}\%)$ when boundary is set as 0.02 and 0.04 respectively}, which are all consistent with the results of the boundary as 0.03 within $1-\sigma$ uncertainties.

Based on the above analysis, we therefore conclude that our conclusions (Fig. \ref{figFAMMR}, Sect. 5) are not (significantly)
affected by the variation of boundary of resonance offset.

\section{Comparison with the results derived from the synthetic in-situ population}
\label{sec.synthetic.insitu}
We analyse the distribution of the period ratios of planet population generated by the Bern model involving the disk migration in the Sect. \ref{sec.PR.theory}. 
We find that ex-situ water-rich sub-Neptunes (mainly Class II) have a larger fraction of planet pairs to be near-mean motion resonance (MMR).
Whereas, for close planets that approximately formed in-situ (Class I), their frequency in MMRs 
is a bit lower but statistically indistinguishable from that of  random distribution.
These results are in agreement with the proposal of the previous studies that convergent disk migrations will lead to more capture in resonance \citep[e.g.][]{2001A&A...374.1092S,2002ApJ...567..596L,2008A&A...483..633P,2021A&A...656A..69E}.

\begin{figure}[!t]
\centering
\includegraphics[width=0.9\linewidth]{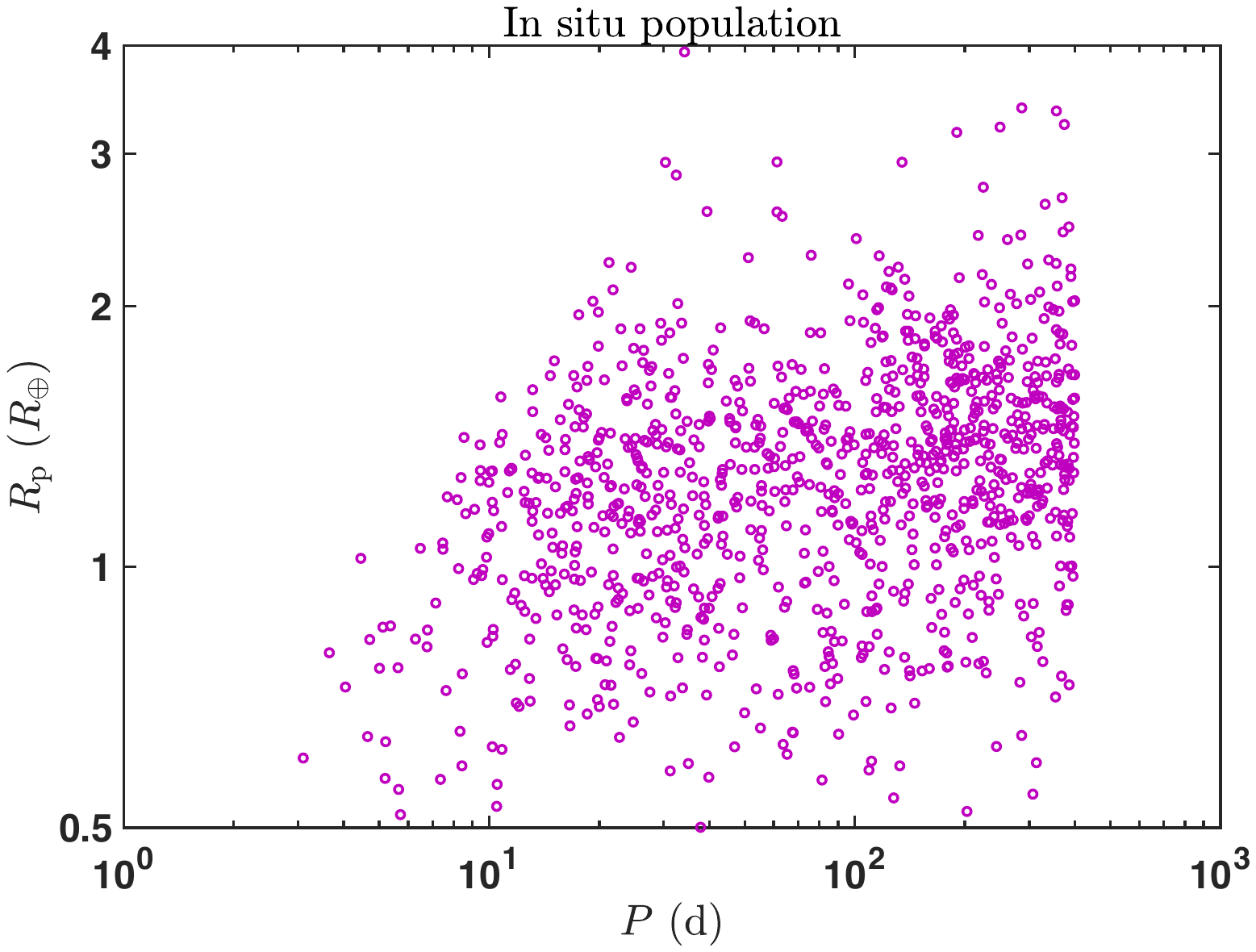}
\caption{Radii as a function of orbital period of planets selected from the synthetic in-situ population without applying the bias of the Kepler survey.
We restricted the sample to planets with radii between $1-4 R_\oplus$ and orbital periods $\le 400$ days in multiple systems.
\label{figradiusperioddigram_insitu}}
\end{figure}

\begin{figure}[!t]
\centering
\includegraphics[width=0.85\linewidth]{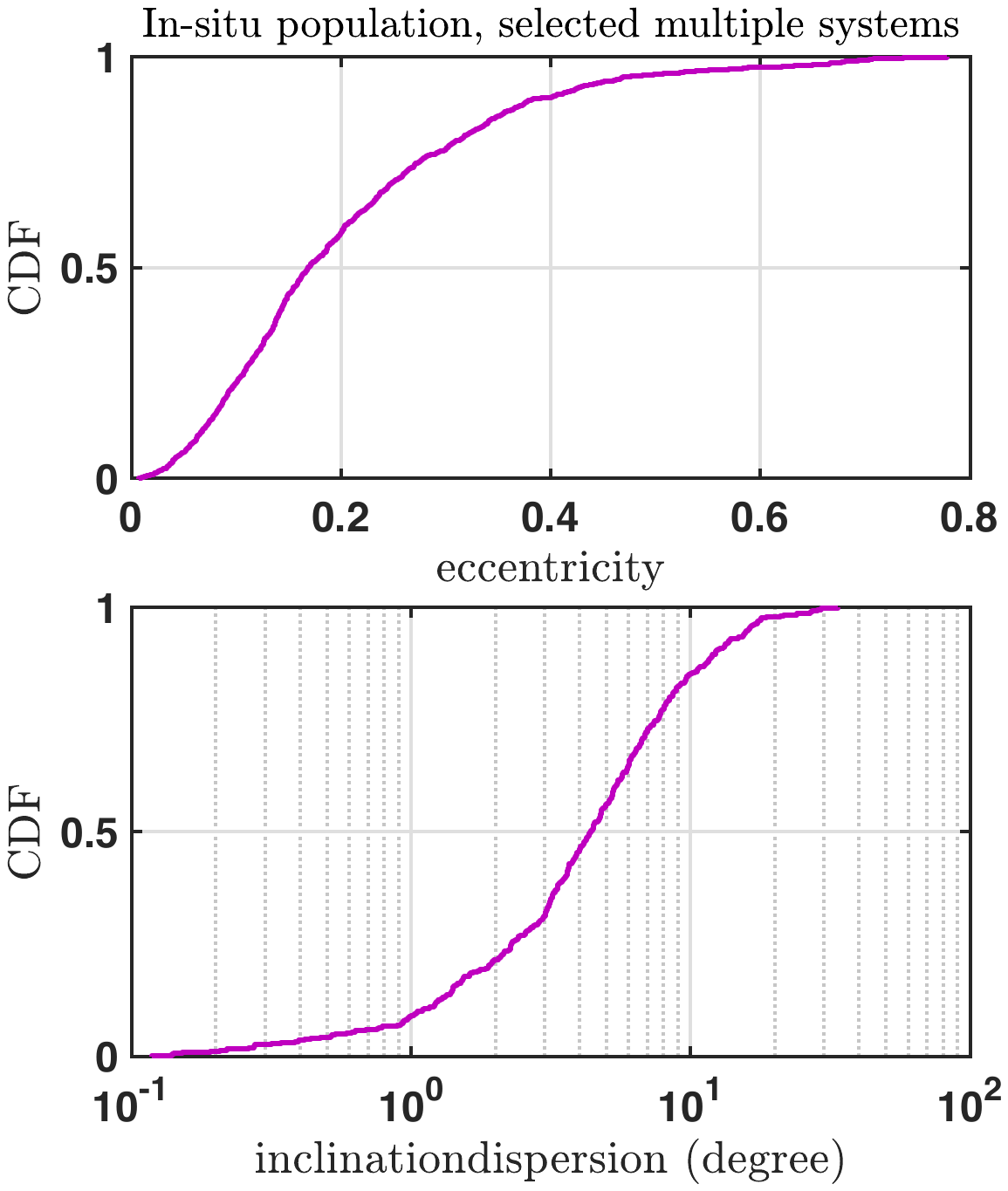}
\caption{The cumulative distributions of the orbital eccentricity (Top) and inclination dispersion (Bottom) for the select synthetic in-situ sample.
\label{figeccinc_insitu}}
\end{figure}

To further verify the above proposal, we also generated a synthetic in-situ planet population using the Bern model without considering disk-planet interactions, i.e. without orbital migration and damping of eccentricities and inclinations (see Appendix A of Emsenhuber et al. 2021b).
With the same criteria described in Sect. 2 of the main text, we select Kepler-like planets ($R_{\rm p} \le 4 R_\oplus$) from the in-situ population at 5 Gyr and only retain systems of two or more planets.
The selected in-situ sample contains 1,031 planets in 419 systems and their radius-period diagram is shown in Fig. \ref{figradiusperioddigram_insitu}. 
Here we do not adopt the KOBE program to apply the Kepler bias because very few systems would be identified as multiple systems after applying the bias.
The main reasons are as follows:
(1). The in-situ population is further away from the host stars and less planets (especially for super-Earths) are detectable;
(2). More importantly, their eccentricities and inclinations have not been damped during the disk phase because all disk-planet interaction were turned off (see Fig. \ref{figeccinc_insitu}).
Multiple planets in such systems are very difficult to detect simultaneously via transits.

Then we calculate the period ratio of adjacent planet pairs in the select in-situ sample.
We also construct a randomly-paired control sample with the same method described in Sect. 3.1.
Since the eccentricities of planets of in-situ population are relatively large, we here calculate the Hill-stability metric with the following equation:
\begin{equation}
    H \equiv \frac{a_{\rm out} (1-e_{\rm out{}})-a_{\rm in} (1+e_{\rm in})}{R_{\rm H}}.
\end{equation}
Then we keep planet pairs as hill-stable with the same criteria described in Sect. 3.2 of the main text.

\begin{figure}[!t]
\centering
\includegraphics[width=0.8\linewidth]{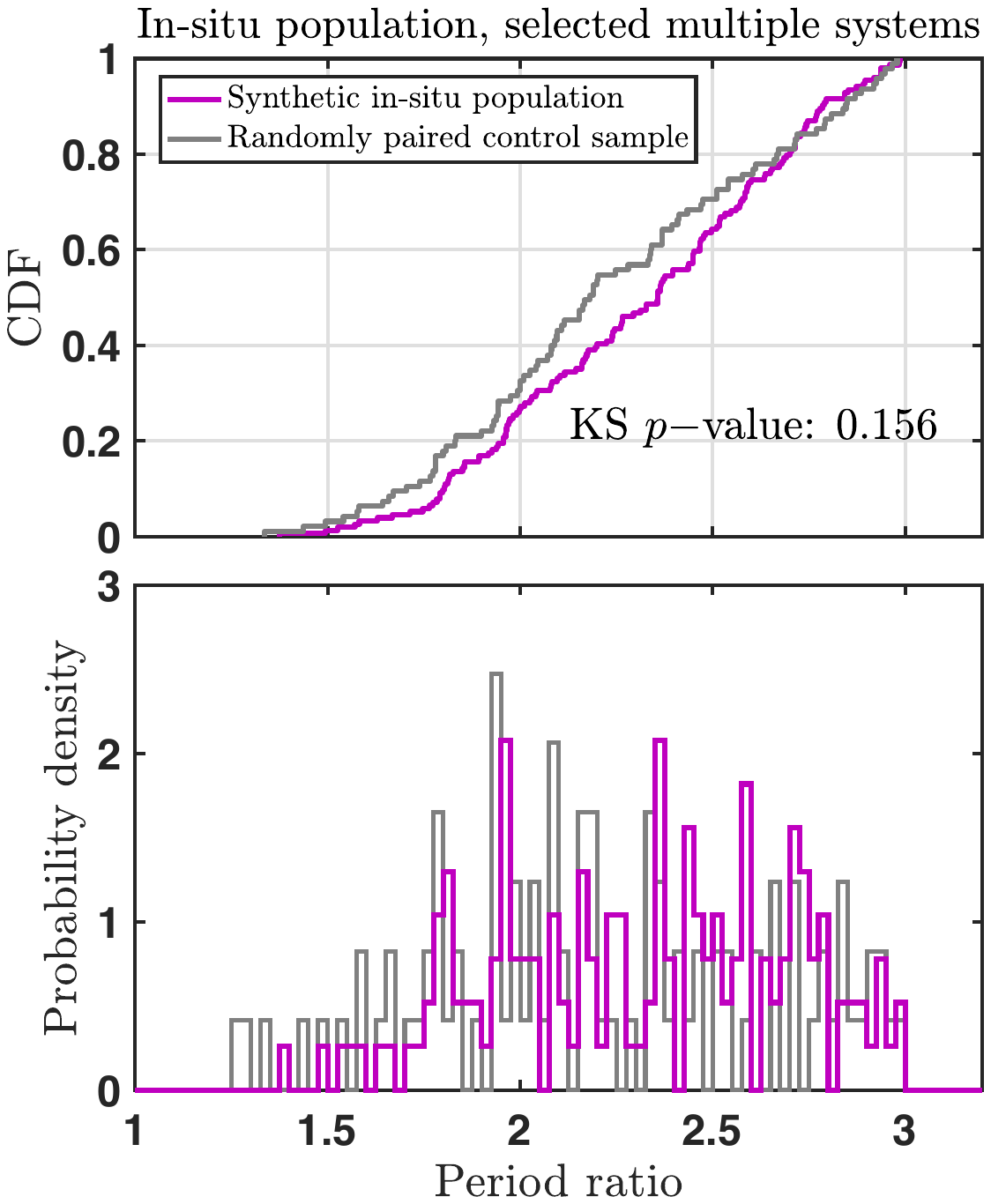}
\caption{The cumulative distributions function (Top) and probability density function (Bottom) of period ratio of adjacent planet pairs derived from the selected in-situ sample (purple) and the corresponding control sample (grey).
In the top-right corner of the top panel, we print the two-sample K-S $p-$ value.
\label{figPR_insitu}}
\end{figure}

Figure \ref{figPR_insitu} shows the cumulative distributions function (Top) and probability density function (Bottom) of the orbital period ratios of planet pairs.
As can be seen, the period ratios of planet pairs in the selected in-situ sample are statistically indistinguishable from the randomly-paired control sample with a KS $p-$value $>0.05$ and show no preference in near-MMR.
We also calculate the normalised fraction in near-MMR and the derived $\hat{F}_{\rm MMR} = 0.9^{+0.4}_{-0.3}$, which is consistent with that of the random distribution.
The above analyses demonstrate that as expected, the synthetic in-situ population behaves in the similar way as a random distribution (but originates still from the same formation model including the N-body interactions but excluding migration and damping).

\end{document}